\documentstyle[12pt,preprint,aps,epsfig,floats, amssymb]{revtex}

\newcommand{\be}{\begin{equation}}
\newcommand{\ee}{\end{equation}}
\newcommand{\bea}{\begin{eqnarray}}
\newcommand{\eea}{\end{eqnarray}}
\newcommand{\ba}{\begin{array}}
\newcommand{\ea}{\end{array}}
\newcommand{\lsim}
{{\;\raise0.3ex\hbox{$<$\kern-0.75em\raise-1.1ex\hbox{$\sim$}}\;}}
\newcommand{\gsim}
{{\;\raise0.3ex\hbox{$>$\kern-0.75em\raise-1.1ex\hbox{$\sim$}}\;}}

\tighten
\begin{document}
\preprint{
\noindent
\begin{minipage}[t]{2in}
\begin{flushleft}
\end{flushleft}
\end{minipage}
\hfill
\begin{minipage}[t]{2in}
\begin{flushright}
DESY 01 - 023\\
February 2001\\
\tt{hep-ph/0103005}\\
\vspace*{.2in}
\end{flushright}
\end{minipage}
}
\draft
\title{\bf  Nonminimal Supersymmetric Standard Model  
with Baryon and Lepton Number Violation}

\author{P. N. Pandita}

\address{Theory Group, Deutsches Elektronen-Synchrotron DESY, 
D-22607 Hamburg, Germany}
\vspace{3mm}
\address{Department of Physics,
North-Eastern Hill University,  Shillong 793 022, India\thanks{Permanent address}}

\maketitle

\begin{abstract}
We carry out a comprehensive analysis of the nonminimal supersymmetric
standard model (NMSSM) with baryon and lepton number violation.
We catalogue the baryon and lepton number violating dimension four
and five operators of the model. We then study
the renormalization group evolution and infrared stable fixed 
points of the Yukawa couplings and the soft 
supersymmetry breaking trilinear couplings of this
model with baryon and lepton number (and R-parity) 
violation involving the heaviest generations.  
We show analytically  that in the Yukawa 
sector of the NMSSM there is only
one infrared stable fixed point. This corresponds
to a non-trivial fixed point for the top-, bottom-quark 
Yukawa couplings and the
$B$ violating coupling $\lambda_{233}''$, and a trivial one for all
other couplings.  All other possible
fixed points are either 
unphysical or unstable in the infra-red region.  
We also carry out an analysis of the renormalization group
equations for the soft supersymmetry breaking trilinear 
couplings, and determine the corresponding fixed points for these
couplings. We then study the quasi-fixed point behaviour, both 
of the third generation Yukawa couplings and the baryon number 
violating coupling, and those of the soft supersymmetry 
breaking trilinear couplings.
From the analysis of the fixed point behaviour, we obtain upper and
lower bounds on the baryon number violating coupling 
$\lambda_{233}''$,
as well as on the soft supersymmetry breaking trilinear couplings.
Our analysis shows that the infrared fixed point behavior
of NMSSM with baryon and lepton number violation is similar to
that of MSSM.
\end{abstract}

\vspace{3mm}



\section{Introduction}

Supersymmetry~\cite{wess} is at present the only known framework in which
the Higgs sector of the Standard Model (SM), so crucial for its
internal consistency, is natural. 
A much favored implementation of
the idea of low energy supersymmetry is the Minimal
Supersymmetric Standard Model (MSSM), obtained by simply
doubling the number of states of the SM, and introducing
a second Higgs doublet (with opposite hypercharge to the
SM Higgs doublet) to generate masses for all the fermions
and to cancel traingle gauge anomalies. The minimal supersymmetric
version of the standard model leads to a successful 
prediction for the ratio of the gauge couplings with a 
gauge unification scale $M_G \simeq 10^{16}$ GeV.  
This has led to the idea that there may
be a stage of unification beyond the SM.  
If so, then it becomes important
to perform the radiative corrections in determining all the dimension
$\leq 4$ terms in the lagrangian.  This can be achieved by using the 
renormalization group equations in finding the values of parameters
at the low scale, given their value at a high scale.
Thus, considerable attention has recently
been focussed on the renormalization group evolution~\cite{schrempp1}
of the various dimensionless Yukawa couplings in the SM and its minimal
supersymmetric extension, the Minimal Supersymmetric Standard Model.
Using the renormalization group evolution, one may attempt to relate the
Yukawa couplings to the gauge couplings via the Pendleton-Ross infra-red
stable fixed point (IRSFP) for the top-quark Yukawa coupling~\cite{pendross},
or via the quasi-fixed point behaviour~\cite{hill}.  The predictive power
of the SM and its supersymmetric extensions may, thus, be enhanced if
the renormalization group (RG) running of the parameters is dominated by
infra-red stable fixed  points (IRSFPs).  
Typically, these fixed points are for ratios like Yukawa couplings
to the gauge coupling, or in the context of supersymmetric models, the 
supersymmetry breaking trilinear $A$-parameter to the gaugino mass, etc.
These ratios do not attain their fixed point values at the weak scale, the
range between the GUT (or Planck) scale and the weak scale being too small
for the ratios to closely approach the fixed point.  Nevertheless, the
couplings may be determined by quasi-fixed point behaviour~\cite{hill}
where the value of the Yukawa coupling at the weak scale is independent of
its value at the GUT scale, provided the Yukawa coupling at the GUT scale
is large.  For the fixed point or the quasi-fixed point scenarios to
be successful, it is necessary that these fixed points  are
stable~\cite{allanach,abel,jack}.  

Since supersymmetry~\cite{wess} requires 
the introduction of superpartners of
all known particles in the SM (in addition to the introduction of at least
two Higgs doublets), which transform in an identical manner under the gauge
group, there are additional Yukawa couplings in supersymmetric models
which violate~\cite{weinberg} baryon number~($B$) or lepton number~($L$).
In MSSM,  a discrete symmetry~\cite{farrar}
called R-parity ($R_p$) is invoked to
eliminate these $B$ and $L$ violating Yukawa couplings.
However, the assumption of $R_p$ conservation at the level of the MSSM
appears to be {\it ad hoc}, since it is not required for the internal
consistency of the model.  Therefore, the study of the renormalization
group evolution of the dimensionless Yukawa couplings in  the MSSM,
including $B$ and $L$ (and $R_p$) violation, deserves serious consideration.
Recently considerable attention has been devoted to this 
question~\cite{ap1,ap2}. It has been shown that the only 
stable infrared fixed point of MSSM with baryon and 
lepton number violation is the one where the 
top-, bottom-quark
Yukawa couplings and the
$B$ violating coupling $\lambda_{233}''$ approach 
a non-trivial fixed point, and 
the $\tau$-Yukawa coupling approaches a trivial fixed point.  
All other possible fixed points are either unphysical
or unstable  in the infrared region.

It is well known that the minimal supersymmetric standard model suffers 
from the so-called $\mu$ problem associated with the bilinear term
connecting the two Higgs doublet superfields in the superpotential.
A simple solution to this problem is to postulate the existence of 
a Higgs singlet superfield, and to couple it to the two Higgs doublets 
in the superpotential via a dimensionless trilinear coupling. When the 
Higgs singlet obtains a vacuum expectation value, a bilinear term
involving the two Higgs doublets is naturally generated~\cite{fayet}.
However, this leads to additional trilinear superpotential couplings
in the model, the so called nonminimal supersymmetric standard model
(NMSSM).  Furthermore, if we do not postulate $R_p$ conservation,
then there is also an additional lepton number
violating superpotential coupling in this model as compared to the 
MSSM with baryon and lepton number violation. 
Because of these additional
trilinear couplings, it is important to study the infra-red fixed point
structure of the NMSSM, and analyze the effect of these additional 
Yukawa couplings on the infrared behavior of the other Yukawa , and 
the baryon- and lepton-number violating couplings, and contrast
it with the situation that obtains in the MSSM. Some preliminary
studies  of the NMSSM with $B$ and $L$ violation
were carried out in~\cite{paulraj}. 
In this paper we carry out a detailed study of the 
renormalization group evolution of the Yukawa couplings of NMSSM, 
including the $B$ and $L$ violating couplings.  
We shall include all the third generation Yukawa couplings, 
as well as the highest generation $B$ and $L$ violating
couplings in our study, and analyze the situation where 
all of them could simultaneously approach infrared fixed points.  
We shall investigate both the true infrared
fixed points, as well as quasi-fixed-points of these couplings.
In particular, we shall carry out a detailed stability analysis of the
infrared fixed points of these couplings.  
Furthermore, corresponding to the $B$ and $L$~(and $R_p$)  violating
Yukawa couplings of the NMSSM, there are soft supersymmetry
breaking trilinear couplings (the $A$ parameters)
whose renormalization group evolution
and infrared fixed point structure has not been studied so far.  
We shall, therefore,  also study the renormalization
group evolution of these soft supersymmetry breaking $A$ parameters, 
including those corresponding to the third generation Yukawa couplings, 
and obtain the simultaneous infrared fixed points for them.

The plan of the paper is as follows.  In Sec. II we describe the model
and write down all the dimension four and five baryon and lepton number
violating couplings in the NMSSM. We then derive  
the renormalization group equations (RGE's) of interest to us,
which include the equations for the dimension four baryon and 
lepton number violating Yukawa couplings.
We then carry out a detailed analytical study of the 
true infrared fixed points of the Yukawa couplings in 
full generality. Within the context of grand unified theories, one is
led to the situation where $B$ and $L$ violating Yukawa couplings 
may be related at the GUT scale, and one may no longer be able to set 
one or the other arbitarily to zero. {\it We, therefore, initially 
include both baryon and lepton number violating couplings
in our RG equations.}  The fixed point analysis of such a system 
of RG equations leads to the crucial result that
the  only stable fixed point is the one with simultaneous non-trivial 
fixed point values for the top- and bottom-quark Yukawa couplings and 
the $B$-violating coupling $\lambda_{233}''$, and a trivial one for 
all other couplings. Thus, non-trivial 
simultaneous fixed points for the $B$ and $L$ violating Yukawa couplings
are ruled out by our analysis of the NMSSM. This is result is analogous
to the corresponding result obtained in  the MSSM~\cite{ap2}.
We then study
the fixed points of the corresponding soft supersymmetry breaking 
trilinear couplings of this model.  In Sec. III
we algebraically study the simultaneous 
quasi-fixed points of all the third 
generation Yukawa couplings of the minimal 
supersymmetric standard model with $B$ violation, 
as well as those of the corresponding soft supersymmetry 
breaking trilinear couplings. Since the quasi-fixed point
limit is formally defined as the Landau pole of the 
Yukawa   coupling at the GUT scale, it provides an upper bound
on the corresponding Yukawa coupling. 
In Sec. IV we present the numerical results 
for the renormalization group evolution and the quasi-fixed points
for the nonminimal supersymmetric standard model with $B$ 
violation.  In Sec. V we summarize  our results and
present the conclusions.

\section{Renormalization Group Equations and Infra-red Fixed Points}

\subsection{Baryon and lepton number violation in NMSSM}

In this section we study the true infra-red fixed points of the 
Yukawa couplings and the $A$ parameters of the NMSSM with $B$ and $L$
violation.  We begin by recalling the basic features of the
NMSSM with baryon and lepton number violation. The superpotential
of the model is written as
\begin{equation}\label{nmssmw}
W = (h_U)_{ab} Q^a_L \overline{U}^b_R H_2
+ (h_D)_{ab} Q^a_L \overline{D}^b_R H_1 + (h_E)_{ab} L^a_L
\overline{E}^b_R H_1 + \lambda N H_1H_2 - \frac{\kappa}{3}N^3.
\end{equation}
where $L,\, Q, \, \overline{E},\, \overline{D},\, \overline{U}$
denote the lepton and quark doublets, and anti-lepon singlet, d-type 
anti-quark singlet and u-type anti-quark singlet,  respectively.
In Eq.~(\ref{nmssmw}), $(h_U)_{ab}$, $(h_D)_{ab}$
and $(h_E)_{ab}$ are the Yukawa coupling matrices, with $a,\, b,\, c$
as the generation indices.
Gauge invariance, supersymmetry and renormalizability allow the 
addition of the following $L$ and $B$ violating terms to the 
superpotential (\ref{nmssmw}):
\begin{eqnarray}
W_L &=& \tilde\lambda_{a} NL_a H_2
+ {1\over 2}\lambda_{abc} L^a_L L^b_L \overline{E}^c_R
+ \lambda'_{abc} L^a_L Q^b_L\overline{D}^c_R,  \label{Lviolating} \\
W_B &=& {1\over 2}\lambda''_{abc} \overline{D}^a_R
\overline{D}^b_R \overline{U}^c_R,   \label{Bviolating} 
\end{eqnarray}
where the notation~\cite{paulraj} is standard. We note 
that there is
a additional $L$-violating term with the dimensionless
Yukawa coupling $\tilde\lambda_a$ in
(\ref{Lviolating}) which does not have an analogue in the MSSM. 
This term can be rotated away into the R-parity
conserving term $\lambda N H_1H_2$ via an $SU(4)$ rotation
between the superfields $H_1$ and $L_a$. However, this rotation
must be performed at some energy scale, and the term is
regenerated through the renormalization group equations.
The Yukawa couplings
$\lambda_{abc}$ and $\lambda''_{abc}$ are antisymmetric in their
first two indices due to $SU(2)_L$ and $SU(3)_C$ group symmetries,
respectively.  

The dimension-4 terms in the supertentials (\ref{Lviolating}) and 
(\ref{Bviolating}) are the most dangerous terms for nucleon decay,
and some of them must be suppressed. This leads to 
constraints~\cite{barbier}
on the different  couplings $\lambda_{abc}, \lambda'_{abc}$,
and $\lambda''_{abc}$, but considerable freedom remains for
the various $B$ and $L$ violating couplings.
Furthermore, there are dimension-5 operators 
which may lead to nucleon decay suppressed by $1/M$, where $M$ is 
some large mass scale at which the $B$ and $L$ violation beyond that of
NMSSM (and MSSM) comes into play. 
Some of these dimension-5 operators may also lead to
unacceptable nucleon decay if their coeffcients are of order unity, and
therefore must be suppressed.
We tabulate here all the dimension-5 operators which are allowed by
the $SU(3)_C \times SU(2)_L\times U(1)_Y$ gauge symmetry and
the particle content of the NMSSM. These are

\bigskip

\begin{eqnarray}
\begin{array}{ll}
{\cal O}_1 = [QQQL]_F,    \hskip 2cm   & {\cal O}_2 = [\bar U \bar U \bar DE]_F,\\
{\cal O}_3 = [QQQH_1]_F,  \hskip 2cm   & {\cal O}_4 = [Q \bar U \bar E H_1]_F,\\
{\cal O}_5 = [LLH_2H_2]_F, \hskip 2cm  & {\cal O}_6 = [LH_1H_2H_2]_F,\\
{\cal O}_7 = [H_2H_2 \bar E^*]_D, \hskip 2cm & {\cal O}_8 = [H_2^*H_1 \bar E]_D,\\
{\cal O}_9 = [Q\bar UL^*]_D, \hskip 2cm & {\cal O}_{10}= [\bar U \bar D^* 
\bar E]_D,\\
{\cal O}_{11} = [LH_2NN]_F, \hskip 2cm & {\cal O}_{12}  = [LL\bar EN]_F,\\
{\cal O}_{13}  = [LQ\bar DN]_F,  \hskip 2cm & {\cal O}_{14}  = [\bar D\bar D
\bar UN]_F,\\
{\cal O}_{15}  = [LH_2N^*]_D,\hskip 2cm& { }  
\end{array}
\label{dim5}
\end{eqnarray}
where we have suppressed the gauge and family indices. We note that the
baryon and lepton number violating operators ${\cal O}_1,........., 
{\cal O}_{10}$ are the same as the corresponding operators in MSSM
and are subject to  constraints similar to those in MSSM, 
whereas the operators ${\cal O}_{11}, ......., {\cal O}_{15}$ are  
additional dimension-5 $B$ and $L$ violating operators 
specific to the NMSSM. We note that
in NMSSM the simutaneous presence of the combinations 
$L Q \bar{D} N$ and $\bar{D}\bar{D}\bar{U}$, 
or $LQ\bar{D}$ and $\bar{D} \bar{D} \bar{U} N$
needs to be forbidden, since these could lead to fast 
proton  decay. We do not consider the dimension-5 operators (\ref{dim5})
any further in this paper, 
and restrict our attention only to the dimension-4 terms in 
(\ref{nmssmw}), (\ref{Lviolating}) and (\ref{Bviolating}), respectively.

Corresponding to the terms in the superpotentials
(\ref{nmssmw}), (\ref{Lviolating}) and (\ref{Bviolating}),
there are the soft supersymmetry breaking trilinear terms 
which can be written as
\begin{eqnarray}
-V_{\rm{soft}} &=&  
\left[(A_U)_{ab}(h_U)_{ab} \tilde{Q}^a_L \tilde{\overline{U}}^b_R H_2
+(A_D)_{ab}(h_D)_{ab} \tilde{Q}^a_L \tilde{\overline{D}}^b_R H_1
\right.
\nonumber \\ 
&+& \left. (A_E)_{ab}(h_E)_{ab} \tilde{L}^a_L \tilde{\overline{E}}^b_R H_1
+ A_{\lambda} \lambda N H_1 H_2 - \frac{A_{\kappa}}{3} {\kappa} N^3 \right] 
\nonumber \\
&+& \left[(A_{\tilde\lambda})_{a} \tilde\lambda_a N \tilde{L}^a_L H_2
+ {1\over 2}(A_\lambda)_{abc}\lambda_{abc} \tilde{L}^a_L \tilde{L}^b_L 
\tilde{\overline{E}}^c_R
+ (A_{\lambda'})_{abc}\lambda'_{abc} \tilde{L}^a_L \tilde{Q}^b_L
\tilde{\overline{D}}^c_R \right] \nonumber \\
&+& \left[ {1\over 2}(A_{\lambda''})_{abc}\lambda''_{abc} 
\tilde{\overline{D}}^a_R
\tilde{\overline{D}}^b_R \tilde{\overline{U}}^c_R \right],  \label{soft}
\end{eqnarray}
where a tilde over a matter chiral superfield 
denotes its scalar component, and
the notation for the scalar component of the Higgs superfield is the same
as that of the corresponding superfield. In addition there are
soft supersymmetry breaking gaugino
mass terms with the masses $M_i$, with $i = 1, 2, 3$
corresponding to the gauge groups $U(1)_Y, \, SU(2)_L, \,$ and 
$SU(3)_C$, respectively.

The third generation Yukawa couplings are the 
dominant couplings in the
superpotential (\ref{nmssmw}). Therefore,
it is natural to retain only the elements 
$(h_U)_{33}\equiv h_t,$~ 
$(h_D)_{33} \equiv h_b,$~  $(h_L)_{33} \equiv h_\tau $~ 
in each of the Yukawa couplings matrices $h_U,\, h_D,
\, h_L$, setting all other elements equal to zero.  Furthermore, there are
39 independent $L$ violating trilinear couplings 
${\tilde\lambda}_{a}$, $\lambda_{abc}$ and
$\lambda'_{abc}$ in (\ref{Lviolating}).  Similarly, there are
9 independent $B$ violating couplings $\lambda''_{abc}$ in 
the baryon number violating superpotential (\ref{Bviolating}).
Thus, we would have to consider 44 coupled nonlinear evolution equations
for the $L$ violating case and 14 coupled nonlinear equations for the $B$
violating case, respectively to study the
renormalization group evolution
of these couplings in the NMSSM.  
It is clear that
there is a need for a radical
simplification of these equations before we can think of studying the
evolution of the Yukawa couplings in the NMSSM 
with  $B$ and $L$ violation.

In order to render the Yukawa couping evolution equations 
tractable, we, therefore,  need
to make certain plausible assumptions.  Motivated by the generational
hierarchy of the conventional Yukawa  couplings, we shall assume that an
analogous hierarchy amongst the different generations of $B$ and $L$
violating couplings exists.  Thus, we shall retain only the 
couplings  $\tilde\lambda_{3}$,  
$\lambda_{233}$, \, $\lambda'_{333}$,\,  $\lambda''_{233}$,
and neglect the rest.  We note that $B$ and $L$
violating couplings
to higher generations evolve more strongly because of larger Higgs
couplings in their evolution equations, and hence could take larger
values than the corresponding couplings to the lighter generations.
We also note that the experimental upper limits are stronger for the
$B$ and $L$ violating couplings in MSSM 
with lower indices~\cite{barbier}.

\subsection{Renormalization group equations}

We are interested in the one-loop renormalization group equations
for the dimensionless trilinear Yukawa couplings in the superpotential
(\ref{nmssmw}), (\ref{Lviolating}) and (\ref{Bviolating}). 
For a general N=1
supersymmetric theory with a trilinear superpotential
term $f_{abc}\Phi^a
\Phi^b\Phi^c$
involving chiral superfields $\Phi^a, ~\Phi^b,
~\Phi^c$, the evolution of the couplings $f_{abc}$ with the 
scale parameter $\mu$
is given by the RGEs~\cite{nfal}
\be
16\pi^2\frac{\partial f_{abc}}{\partial\ln\mu} = \gamma_a^e f_{ebc}
+\gamma_b^e f_{aec} + \gamma_c^e f_{abe},
\label{rg1}
\ee
where $\gamma_a^e$ are the elements of the anomalous dimension matrix,
and sum over repeated indices is understood. The anomalous dimensions
are given by
\be
\gamma_a^e = \frac{1}{2}\sum_{b,c} f_{abc} f^{ebc}
- 2\delta_a^e g_A^2 C_a^A,
\label{ganam}
\ee
to one loop order. The sum over A represents a sum over all dominant
gauge couplings, and $C_a^A$ is the quadratic Casimir of the
representation of $\Phi^a$ under the gauge group with 
coupling $g_A$:
\be
(T_R^A~T_R^A)_a^b = C_R^A~\delta_a^b.
\label{trace}
\ee
Here, $T_R^A$ is a matrix in the R representation for the group labelled
by A. Pictorially, the RG evolution of the trilinear coupling can be
described as an insertion of the anomalous dimension correction on each
external leg. We have calculated the anomalous 
dimension for the various superfields for the NMSSM 
with baryon- and lepton-number violating couplings. 
These are summarized in Table~\ref{table1}.
\begin{table}[htb]
\begin{footnotesize}
\begin{center}
\vskip6pt
\renewcommand\arraystretch{1.2}
\begin{tabular}{|lccc|}
\multicolumn{1}{|c}{$\Phi^{a,b}$} & NMSSM & L violation
& B violation\\
\hline
$\hat N, \hat N$ & $4\lambda^2 + 4k^2$ & $\lambda^i \lambda_i$ & $-$ \\
$\hat L_i, \hat H_1$ & $-$ & $\lambda^{iab}(h_E)_{ab}+3\lambda^{'iab}(h_D)_{ab}
+\lambda\lambda^i$ & $-$ \\
$\hat L_{i,j}$ & $h_L h_L^{\dagger} - \frac{3}{2}g_2^2 -\frac{3}{10}g_1^2$ &
$\lambda_{iab} \lambda^{jab} + 3\lambda_{iab}^{'}\lambda^{'jab}
+\lambda_i\lambda^i\delta_j^i$ & $-$ \\
$\hat E_{i,j}^c$ & $2h_E^{\dagger}h_E - \frac{6}{5}g_1^2$ & $\lambda^{abi}
\lambda_{abj}$ & $-$ \\
$\hat D_{i,j}^c$ & $2h_D^{\dagger}h_D- \frac{8}{3}g_3^2-\frac{2}{15}g_1^2$ &
$2\lambda^{'abi}\lambda_{abj}^{'}$ &
$2\lambda^{''iab}\lambda_{jab}^{''}$\\
$\hat U_{i,j}^c$ & $2h_U^{\dagger}h_U- \frac{8}{3}g_3^2-\frac{8}{15}g_1^2$ &
$-$ & $\lambda^{''abi}\lambda_{abj}^{''}$ \\
$\hat Q_{i,j}$ & $h_U h_U^{\dagger} + h_D h_D^{\dagger} - \frac{8}{3}g_3^2
- \frac{3}{2}g_2^2 - \frac{1}{30}g_1^2$ & $\lambda^{'}_{aib}
\lambda^{'ajb}$ & $-$\\
$\hat H_1, \hat H_1$ & $Tr(h_E h_E^{\dagger} + 3h_D h_D^{\dagger})+\lambda^2
-\frac{3}{2}g_2^2 -\frac{3}{10}g_1^2$ & $-$ & $-$\\
$\hat H_2, \hat H_2$ & $3Tr(h_U h_U^{\dagger}) + \lambda^2-\frac{3}{2}g_2^2
-\frac{3}{10}g_1^2$ & $\lambda_i\lambda^i$ & $-$\\
\end{tabular}
\end{center}
\caption{The anomalous dimensions $\gamma_{\Phi^b}^{\Phi^a}$ in the
non-minimal supersymmetric standard model with lepton and baryon number
violating couplings. Here $a, b$  are flavour indices.}
\label{table1}
\end{footnotesize}
\end{table}
The renormalization group equation for the Yukawa couplings $h_U,
h_D, h_E$ of the superpotential (\ref{nmssmw}) are obtained from
(\ref{rg1}) with the index c belonging to a Higgs field. The general
form of the RGEs are
\bea
16\pi^2 \frac{\partial}{\partial\ln\mu}(h_U)_{ab} &=& (h_U)_{ib}
\gamma_{Q_a}^{Q_i} + (h_U)_{ai}\gamma_{\bar U_b}^{\bar U_i}
+(h_U)_{ab}\gamma_{H_2}^{H_2},
\label{HU}
\\
16\pi^2 \frac{\partial}{\partial\ln\mu}(h_D)_{ab} &=& (h_D)_{ib}
\gamma_{Q_a}^{Q_i} + (h_D)_{ai}\gamma_{\bar D_i}^{\bar D_i}
+(D)_{ab}\gamma_{H_1}^{H_1} + \lambda_{iab}^{'}\gamma_{H_1}^{L_i},
\label{HD}
\\
16\pi^2 \frac{\partial}{\partial\ln\mu}(h_L)_{ab} &=& (h_L)_{ib}
\gamma_{L_a}^{L_i} + (h_L)_{ai}\gamma_{\bar E_b}^{\bar E_i}
+(h_L)_{ab}\gamma_{H_1}^{H_1} + \lambda_{iab}\gamma_{H_1}^{L_i}.
\label{HL}
\eea
The evolution of gauge couplings $g_i$ ($i=1,2,3$~~denoting the $U(1)_Y$,
$SU(2)_L$ and $SU(3)_C$ gauge groups, with GUT normalization for
$U(1)_Y$ gauge group) in NMSSM with $B$- and $L$-violation
is same as in MSSM, as this evolution is unaffected at the one-loop
level by the presence of a chiral singlet superfield, 
or B- and L-violating couplings.  These evolution equations are
\be
16\pi^2 \frac{dg_i}{d\ln\mu} = b_i g_i^3,~~~i = 1,2,3,
\label{gauge1}
\ee
where $b_i$ are the beta functions for the respective gauge couplings
with $b_1=33/5,~b_2=1,~b_3=-3$. The corresponding one-loop 
renormalization group equations  for the gaugino masses $M_i$, i = 1, 2, 3
can be written as
\be
16\pi^2 \frac{dM_i}{d\ln\mu} = 2 g_i^2 b_i M_i,~~~i = 1,2,3.
\label{gaugino1}
\ee

Retaining only the third generation Yukawa couplings and the highest 
generation baryon and lepton number violating  couplings, the
RGEs for the Yukawa couplings and the R-parity violating
couplings in the NMSSM can be written as~\cite{PP21}
\bea
\frac{dh_t}{d\ln\mu}&=&\frac{h_t}{16\pi^2}\left[6h_t^2+h_b^2+\lambda^2
+\tilde\lambda_3^2+\lambda_{333}^{'2}+2\lambda_{233}^{''2}\right. \nonumber \\
&&\left. -\left(\frac{16}{3}g_3^2+3g_2^2+\frac{13}{15}g_1^2\right)\right], 
\label{HT}\\
\frac{dh_b}{d\ln\mu}&=&\frac{1}{16\pi^2}\left[\left(h_t^2+6h_b^2+h_{\tau}^2
+ \lambda^2 + 6\lambda_{333}^{'2}+2\lambda_{233}^{''2}\right)h_b
\right.+\lambda\tilde\lambda_3\lambda_{333}^{'}\nonumber\\
&&\left.-\left(\frac{16}{3}g_3^2+3g_2^2
+\frac{7}{15}g_1^2\right)h_b\right], 
\label{HB}\\
\frac{dh_{\tau}}{d\ln\mu}&=&\frac{h_{\tau}}{16\pi^2}\left[3h_b^2+4h_{\tau}^2
+\lambda^2+\tilde\lambda_3^2+4\lambda_{233}^2+3\lambda_{333}^{'2}
\right. \nonumber \\
&& \left. -\left(3g_2^2+\frac{9}{5}g_1^2\right)\right],
\label{AHTA}\\
\frac{d\lambda}{d\ln\mu}&=&\frac{1}{16\pi^2}\left[\left(3h_t^2+3h_b^2
+h_{\tau}^2+4\lambda^2+2{\kappa}^2+4\tilde\lambda_3^2\right)\lambda
+3h_b\tilde\lambda_3 \lambda_{333}^{'}\right.\nonumber\\
&&\left.-\left(3g_2^2+\frac{3}{5}g_1^2\right)\lambda\right], 
\label{DLA}\\
\frac{d{\kappa}}{d\ln\mu}&=&\frac{{\kappa}}{16\pi^2}\left[6\lambda^2
+6{\kappa}^2 +6\tilde\lambda_3^2\right], 
\label{DKA}\\
\frac{d\tilde\lambda_3}{d\ln\mu}&=&\frac{1}{16\pi^2}
\left[\left(3h_t^2+h_{\tau}^2
+4\lambda^2+2{\kappa}^2+4\tilde\lambda_3^2+\lambda_{233}^2+3\lambda_{333}^{'2}
\right)\tilde\lambda_3\right. \nonumber\\
&&\left.+3h_b\lambda\lambda_{333}^{'}
-\left(3g_2^2+ \frac{3}{5}g_1^2\right)\tilde\lambda_3\right], 
\label{DL}\\
\frac{d\lambda_{233}}{d\ln\mu}&=&\frac{\lambda_{233}}{16\pi^2}\left[4h_{\tau}^2
+\tilde\lambda_3^2+4\lambda_{233}^2+3\lambda_{333}^{'2}-\left(3g_2^2
+\frac{9}{5}g_1^2\right)\right], 
\label{DLT}\\
\frac{d\lambda_{333}^{'}}{d\ln\mu}&=&\frac{1}{16\pi^2}\left[\left(h_t^2
+6h_b^2+h_{\tau}^2+\tilde\lambda_3^2+\lambda_{233}^2+6\lambda_{333}^{'2}
+2\lambda_{233}^{''}\right)\lambda_{333}^{'}\right.\nonumber\\
&&\left.+h_b\lambda\tilde\lambda_3
-\left(\frac{16}{3}g_3^2+3g_2^2
+\frac{7}{15}g_1^2\right)\lambda_{333}^{'}\right],
\label{DLP}\\
\frac{d\lambda_{233}^{''}}{d\ln\mu}&=&\frac{\lambda_{233}^{''}}{16\pi^2}
\left[\left(2h_t^2+2h_b^2+2\lambda_{333}^{'2}+6\lambda_{233}^{''2}
\right)-\left(8g_3^2+\frac{4}{5}g_1^2\right)\right].
\label{DLD}
\eea
We note that since the difference between the one-loop 
and two-loop results~\cite{codkaz1} for the infrared fixed points 
in MSSM is less than $10\%$, we shall use one-loop 
renormalization group equations in the  study of infrared
fixed points in the NMSSM in this paper.

We now come to the evolution equations for the soft supersymmetry
breaking trilinear parameters in the potential 
(\ref{soft}).  The one-loop RGEs 
for these parameters can be deduced from the general expressions in
ref.~\cite{rges}.  In this paper
we shall assume the same kind of generational
hierarchy for these trilinear parameters as was assumed for the
corresponding Yukawa couplings.  Thus,  we shall consider, besides
$A_{\lambda}$ and $A_{\kappa}$,  only the 
highest generation trilinear coulings 
$(A_U)_{33}\equiv A_t$,  $(A_D)_{33}\equiv A_b$, $(A_L)_{33}\equiv A_\tau$,
$(A_{\tilde\lambda})_3\equiv A_{\tilde\lambda_3}$
$(A_\lambda)_{233}\equiv A_{\lambda_{233}}$, 
$(A_{\lambda'})_{333}\equiv A_{\lambda'_{333}}$,
$(A_{\lambda''})_{233}\equiv A_{\lambda''_{233}}$, setting all other elements
equal to zero.  With this assumption the RGEs for the soft supersymmetry 
breaking trilinear parameters can be written as
\bea
{dA_t\over d\ln \mu} & = &{1\over 8 \pi^2}
\left(6 A_t h_t^2 + A_b h_b^2 
+ A_{\lambda} \lambda^2 + A_{\tilde\lambda_3}\tilde\lambda^2_3
+ A_{\lambda'_{333}} \lambda'^2_{333} 
+ 2 A_{\lambda''_{233}} \lambda''^2_{233} \right. \nonumber\\
&&\left. - {16\over 3}M_3 g_3^2 - 3 M_2 g_2^2 - 
{13\over 15} M_1 g_1^2 \right),  \label{rge7}\\
{dA_b\over d\ln \mu} & = & {1\over 8\pi^2}
\left(A_t h_t^2 + 6 A_b h_b^2 + A_\tau h_\tau^2 + A_{\lambda} \lambda^2
+ \frac{3}{2} A_b \lambda'^2_{333} 
+ \frac{9}{2} A_{\lambda'_{333}}\lambda'^2_{333} \right.  \nonumber\\
&& \left. + 2 A_{\lambda''_{233}}\lambda''^2_{233}
- \frac{A_b\lambda\tilde\lambda_3\lambda'_{333}}{2 h_b}
+  \frac{A_{\lambda}\lambda\tilde\lambda_3\lambda'_{333}}{h_b}
+ \frac{A_{\lambda'_{333}}\lambda\tilde\lambda_3\lambda'_{333}}{2 h_b}
\right. \nonumber\\
&& \left. - {16\over 3}M_3 g_3^2 - 3 M_2 g_2^2 - {7\over 15}M_1 g_1^2 \right),  
\label{rge8}\\
{dA_\tau\over d\ln \mu} & = & {1\over 8 \pi^2}
\left( 3 A_b h_b^2 + 4 A_\tau h_\tau^2 
+ A_\lambda \lambda^2 + A_{\tilde\lambda_3} \tilde\lambda^2_3
+ \frac {A_{\tau}\lambda_{233}^2}{2}
+ \frac {7 A_{\lambda_{233}}\lambda_{233}^2}{2}\right. \nonumber\\
&& \left. + 3 A_{\lambda'_{333}}\lambda'^2_{333}
 - 3 M_2 g_2^2 - {9\over 5}M_1 g_1^2 \right),   \label{rge9}\\
{d A_\lambda\over d\ln \mu} & = & {1\over 8 \pi^2}
\left(3 A_t h_t^2 + 3 A_b h_b^2 +  A_\tau h_\tau^2
+ 4 A_\lambda\lambda^2 
+ \frac {A_\lambda \tilde\lambda_3^2}{2}
+ 2 A_{\kappa} {\kappa}^2 \right. \nonumber\\
&& \left.+\frac {7 A_{\tilde\lambda_3} \tilde \lambda_3^2}{2}
+\frac{3 A_b h_b \tilde\lambda_3\lambda'_{333}}{\lambda}
-\frac{3 A_\lambda h_b \tilde\lambda_3\lambda'_{333}}{2 \lambda}
+\frac{3 A_{\tilde\lambda_3} h_b \tilde\lambda_3\lambda'_{333}}{2 \lambda}
\right. \nonumber\\
&&\left. - 3 M_2 g_2^2 - {3\over 5}M_1 g_1^2 \right),   \label{rge10}\\
{d A_{\kappa}\over d\ln \mu} & = & {6\over 8 \pi^2}
\left(A_\lambda \lambda^2  +  A_{\kappa} {\kappa}^2 
+ A_{\tilde\lambda_3}\tilde\lambda_3^2 \right),   \label{rge11}\\
{dA_{\tilde \lambda_3}\over d\ln \mu} & = & {1\over 8 \pi^2}
\left(3 A_t h_t^2 + A_\tau h_\tau^2 + \frac{7 A_\lambda\lambda^2}{2}
+2 A_{\kappa} {\kappa}^2 +\frac{A_{\tilde\lambda_3} \lambda^2}{2}
+ 4 A_{\tilde\lambda_3} \tilde\lambda_3^2 \right. \nonumber\\
&& \left. + A_{\lambda_{233}}\lambda^2_{233} 
+ 3 A_{\lambda'_{333}}\lambda'^2_{333}
+ \frac{3 A_\lambda h_b \lambda \lambda'_{333}}{2 \tilde\lambda_3}
- \frac{3 A_{\tilde\lambda_3} h_b \lambda \lambda'_{333}}{2 \tilde\lambda_3} 
\right. \nonumber\\
&&\left. 
+  \frac{3 A_{\lambda'_{333}} h_b \lambda \lambda'_{333}}{\tilde\lambda_3}
- 3 M_2 g_2^2 - {3\over 5}M_1 g_1^2 \right),
\label{rge12}\\
{dA_{\lambda_{233}}\over d\ln \mu} & = & {1\over 8 \pi^2}
\left(\frac{7 A_\tau h_\tau^2}{2} + A_{\tilde\lambda_3} \tilde\lambda_3^2 
+ \frac{A_{\lambda_{233}} h_{\tau}^2}{2}
+ 4 A_{\lambda_{233}}\lambda_{233}^2
+ 3 A_{\lambda'_{233}}\lambda'^2_{333} 
\right. \nonumber \\
&&\left. - 3 M_2 g_2^2 - {9\over 5}M_1 g_1^2 \right),
\label{rge13}\\
{dA_{\lambda'_{333}}\over d\ln \mu} & = & {1\over 8 \pi^2}
\left( A_t h_t^2 + \frac{9A_b h_b^2}{2} + A_\tau h_\tau^2 
+ A_{\tilde\lambda_3} \tilde\lambda_3^2
+ A_{\lambda_{233}}\lambda_{233}^2 \right. \nonumber\\
&& \left. + \frac{3 A_{\lambda'_{333}} h_b^2}{2}
+ 6 A_{\lambda'_{333}}\lambda'^2_{333} 
+ 2 A_{\lambda''_{233}}\lambda''^2_{233} 
+\frac{A_b h_b \lambda \tilde\lambda_3}{2\lambda'_{333}}
+\frac{A_{\tilde\lambda_3} h_b \lambda \tilde\lambda_3}{\lambda'_{333}}
\right. \nonumber \\
&& \left. 
- \frac{A_{\lambda'_{333}} h_b \lambda \tilde\lambda_3}{2\lambda'_{333}}
- {16\over 3} M_3 g_3^2 - 3 M_2 g_2^2 - {7\over 15}M_1 g_1^2 \right), 
\label{rge14}\\
{dA_{\lambda''_{233}}\over d\ln \mu} & = & {1\over 8 \pi^2}
\left(2 A_t h_t^2 + 2 A_b h_b^2 + 2 A_{\lambda'}
\lambda'^2_{333} + 6 A_{\lambda''_{233}}\lambda''^2_{233}
\right. \nonumber \\
&& \left. - 8 M_3 g_3^2  - {4\over 5}M_1 g_1^2 \right).  \label{rge15}
\eea

Given the evolution equations (\ref{HT}) - (\ref{DLD}) for the
Yukawa couplings and the  evolution
equations (\ref{rge7}) - (\ref{rge15}) for the
$A$ parameters, we are now ready to study the RG evolution
and infra-red fixed
points of the NMSSM with $B$ and $L$ violation.

\subsection{Infrared fixed points for Yukawa couplings}
In this section we consider the infrared fixed points for the 
Yukawa couplings and the baryon and lepton number violating couplings
of the NMSSM.
The infra-red fixed
points~\cite{paulraj} of the renormalization group equations
(\ref{HT}) - (\ref{DLD}) have been studied
in the limit of ignoring the $\tau$-Yukawa coupling $h_\tau$, and
by considering either the baryon number violating Yukawa coupling
$\lambda''_{233}$, or the lepton number violating
Yukawa couplings $\tilde\lambda_3$, $\lambda_{233}$
and $\lambda'_{333}$.  In the analysis that follows, we shall consider
the evolution equations for $h_t$ and $h_b$
together with the evolution equation
for $h_{\tau}$.  Furthermore, we shall also entertain the
possibility of simultaneous presence of $B$ and $L$ violating couplings in the
renormalization group equations
(\ref{HT}) - (\ref{DLD}).  
{\it We do this in order to investigate as to whether such a 
system of equations
does have acceptable infrared fixed points.}  However, 
without loss of generality, we shall assume that there is a hierarchy of lepton
number violating couplings, and consider only one lepton number violating
coupling,
together with the baryon number violating coupling $\lambda^{''}_{233}$, at
a time. Thus, we shall consider three different cases, i.e., we shall take
$\tilde\lambda_3 \gg \lambda_{233}, \lambda'_{333}$, 
or $\lambda_{233} \gg \tilde\lambda_3, \lambda'_{333}$,
or $\lambda'_{333} \gg \tilde\lambda_3, \lambda_{233}$. In order to
study the infrared fixed points, we define the following ratios of the 
dimensionless trilinear couplings and the $SU(3)_C$ gauge coupling:
\bea
R_t&=&\frac{h_t^2}{g_3^2},~~R_b=\frac{h_b^2}{g_3^2},~~R_{\tau}=
\frac{h_{\tau}^2}{g_3^2},~~R_{\lambda}=\frac{\lambda^2}{g_3^2},~~
R_{\kappa}=\frac{\kappa^2}{g_3^2},
\label{RY}\\
\tilde R_3&=&\frac{\tilde\lambda_3^2}{g_3^2},~~
R=\frac{\lambda_{233}^2}{g_3^2},~~
R^{'}=\frac{\lambda_{333}^{'}}{g_3^2}, ~~ 
R^{''}=\frac{\lambda_{233}^{''2}}{g_3^2}.
\label{RBL}
\eea
We shall first consider the RG evolution
of Yukawa couplings in the superpotential (\ref{nmssmw}), 
with $B$ violation arising from 
(\ref{Bviolating}), and  $L$ violation arising from the
first term in (\ref{Lviolating}).
With the definitions (\ref{RY}) and (\ref{RBL}), and retaining
only the $SU(3)_C$ gauge coupling, the one-loop renormalization
group equations for $h_t,~h_b,~h_{\tau},~\lambda,~k~$, 
and $\tilde\lambda_3, \lambda_{233}^{''}$ 
can be written in the form 
\bea
\frac{dR_t}{d(-\ln\mu^2)}&=&\tilde\alpha_3 R_t\left[\left(\frac{16}{3}
+b_3\right) - 6R_t - R_b - R_{\lambda} - \tilde R_3 
- 2R^{''}\right],
\label{RT}\\
\frac{dR_b}{d(-\ln\mu^2)}&=&\tilde\alpha_3 R_b\left[\left(\frac{16}{3}
+b_3\right) - R_t - 6R_b - R_{\tau} 
- R_{\lambda} - 2 R^{''}\right],
\label{RB}\\
\frac{dR_{\tau}}{d(-\ln\mu^2)}&=&\tilde\alpha_3 R_{\tau}
[b_3-3R_b-4R_{\tau}-R_\lambda - \tilde R_3],
\label{RTAU}\\
\frac{dR_{\lambda}}{d(-\ln\mu^2)}&=&\tilde\alpha_3 R_{\lambda}
[b_3 -3 R_t - 3R_b - R_{\tau} - 4R_{\lambda}-2R_{\kappa} - 4\tilde R_3],
\label{RLAM}\\
\frac{dR_{\kappa}}{d(-\ln\mu^2)}&=&\tilde\alpha_3 R_{\kappa}[b_3
- 6R_{\lambda} - 6R_{\kappa} - 6 \tilde R_3],
\label{RKAP}\\
\frac{d\tilde R_3}{d(-\ln\mu^2)}&=&\tilde\alpha_3 \tilde R_3[b_3
- 3R_t - R_{\tau} - 4R_{\lambda} - 2R_{\kappa} - 4\tilde R_3] 
\label{RLTIL}\\
\frac{dR^{''}}{d(-\ln\mu^2)}&=&\tilde\alpha_3 R^{''}
[(8+b_3) - 2R_t - 2R_b - 6R^{''}],
\label{RLDP}
\eea
where $b_3=-3$ is the beta function for $g_3$ in NMSSM (or MSSM),
and $\tilde\alpha_3=g_3^2/(16\pi^2)$. Choosing the basis of
the ratios as
$R_i=(R_t, R_b, R_{\tau}, R_{\lambda}, R_{\kappa}, R^{''}, \tilde R_3)$, 
we can rewrite the
RG equations (\ref{RT}) - (\ref{RLDP}) in the form $(t=-\ln\mu^2)$
\be
\frac{dR_i}{dt}=\tilde\alpha_3 R_i[(r_i+b_3)-\sum_j S_{ij} R_j],
\label{RI}
\ee
where $r_i=\sum_{k} 2C_R$, $C_R$ is the QCD quadratic Casimir for the various
fields $(C_Q=C_{U^c}=C_{D^c}=4/3)$ and the sum is over the representation
of the three fields associated with the trilinear coupling that
enters the definition of $R_i$, and $S$ is a matrix whose entries are the
numerical coefficients (the wave function anomalous dimensions) 
of $R_i$'s in the evolution equations (\ref{RT}) - (\ref{RLDP}). 
A fixed point is,  then,  reached
when the right hand side of Eq.~(\ref{RI}) is
0 for all $i$.  If we were to write the fixed point solutions as
$R_i^*$, then there are two fixed point values for each coupling:
$R_i^*=0$, or
\begin{equation}
\left[\left(r_i+b_3\right) -\sum_j S_{ij} R_j^* \right]=0.
\end{equation}
It follows that the non-trivial fixed point solution is
\begin{equation}\label{RISTAR}
R_i^*=\sum_j (S^{-1})_{ij} (r_j+b_3).
\end{equation}
The anomalous dimension matrix S that enters Eq.~(\ref{RISTAR}),  
which we denote by $S_{BL1}$ in this case,  is easily  seen to be
\begin{equation}\label{sbl1}
S_{BL1}=\left[
\begin{array}{c c c c c c c}
6 & 1 & 0 & 1 & 0 & 2 & 1 \\
1 & 6 & 1 & 1 & 0 & 2 & 0 \\
0 & 3 & 4 & 1 & 0 & 0 & 1 \\
3 & 3 & 1 & 4 & 2 & 0 & 4 \\
0 & 0 & 0 & 6 & 6 & 0 & 6 \\
2 & 2 & 0 & 0 & 0 & 6 & 0\\
3 & 0 & 1 & 4 & 2 & 0 & 4\\
\end{array}
\right].
\end{equation}
Inverting the matrix  (\ref{sbl1}) and substituting in
Eq.(\ref{RISTAR}), we get
the following fixed point solution:
\begin{eqnarray}
R_t^*&=&{29\over 76}, \, \, \, R_b^*= 0, \, \, \, 
R_{\tau}^* = -{31 \over 76}, \, \, \, 
R_{\lambda}^* = {18 \over 19}, \, \, \, R_{\kappa}^* = {33 \over 38},  
\nonumber \\
R^{''*} &=& {161 \over 228}, \, \, \,
\tilde R_3^* = -{44\over 19}. \, \, \, 
\end{eqnarray}
We note that $R^*_\tau, \tilde R_3^* < 0$, and, therefore, this fixed point 
solution is physically unacceptable.  
We  conclude  that a simultaneous fixed
point for the $B$ and $L$  violating couplings $\lambda''_{233}$
and $\tilde \lambda_3$,  and the
Yukawa couplings $h_\tau, \, h_b, \, h_t, \lambda, \kappa,$ does
not exist.

Next we consider the RG evolution
of Yukawa couplings in the superpotential (\ref{nmssmw}),
with $B$ violation arising from 
(\ref{Bviolating}), and  $L$ violation arising from the
second term in (\ref{Lviolating}).
The one-loop renormalization
group equations for $h_t,~h_b,~h_{\tau},~\lambda,~\kappa~$,
and $\lambda_{233}$ and $\lambda_{233}^{''}$
can be written in the form (\ref{RI}), with the anomalous dimension
matrix $S$ given by
\begin{equation}\label{sbl2}
S_{BL2}=\left[
\begin{array}{c c c c c c c}
6 & 1 & 0 & 1 & 0 & 0 & 2 \\
1 & 6 & 1 & 1 & 0 & 0 & 2 \\
0 & 3 & 4 & 1 & 0 & 4 & 0 \\
3 & 3 & 1 & 4 & 2 & 0 & 0 \\
0 & 0 & 0 & 6 & 6 & 0 & 0 \\
0 & 0 & 4 & 0 & 0 & 4 & 0 \\
2 & 2 & 0 & 0 & 0 & 0 & 6 \\
\end{array}
\right],
\end{equation}
where the ordering of the ratios is 
$R_i=(R_t, R_b, R_{\tau}, R_{\lambda}, R_{\kappa}, R, R^{''})$.
From (\ref{sbl2}) and (\ref{RISTAR}) we get the fixed point values
\begin{eqnarray}
R_t^*&=&{1\over 2}, \, \, \, R_b^*= {3 \over 4}, \, \, \,
R_{\tau}^* = -{5 \over 4}, \, \, \,
R_{\lambda}^* = -{9 \over 4}, \, \, \, R_{\kappa}^* = { 7 \over 4},  
\nonumber \\
R^* &=& {1 \over 2}, \, \, \,
R^{''*} = {5 \over 12}. \, \, \,
\end{eqnarray}
Since $R_{\tau}^*, R_{\lambda}^* < 0$, this fixed point must also be rejected 
as being unphysical.
Thus, a simultaneous fixed
point for the $B$ and $L$  violating couplings $\lambda''_{233}$
and $\lambda_{233}$,  and the
Yukawa couplings $h_\tau, \, h_b, \, h_t, \lambda, \kappa, $ does
not exist.

Finally,  we consider the RG evolution
of Yukawa couplings in the superpotential (\ref{nmssmw}),
with $B$ violation arising
from (\ref{Bviolating}), and  $L$ violation arising from the
third  term in (\ref{Lviolating}).
The one-loop renormalization
group equations for $h_t,~h_b,~h_{\tau},~\lambda,~\kappa~$,
and $\lambda^{'}_{333}$ and $\lambda_{233}^{''}$
can again be written in the form (\ref{RI}), with the anomalous dimension
matrix $S$ given by\\

\begin{equation}\label{sbl3}
S_{BL3}=\left[
\begin{array}{c c c c c c c}
6 & 1 & 0 & 1 & 0 & 1 & 2 \\
1 & 6 & 1 & 1 & 0 & 6 & 2 \\
0 & 3 & 4 & 1 & 0 & 3 & 0 \\
3 & 3 & 1 & 4 & 2 & 0 & 0 \\
0 & 0 & 0 & 6 & 6 & 0 & 0 \\
1 & 6 & 1 & 0 & 0 & 6 & 2 \\
2 & 2 & 0 & 0 & 0 & 2 & 6 \\
\end{array}
\right],
\end{equation}
where the ordering of the ratios is  
$R_i=(R_t, R_b, R_{\tau}, R_{\lambda}, R_{\kappa}, R^{'}, R^{''})$.
This results in the fixed point values
\begin{eqnarray}
R_t^*&=&{31\over 292}, \, \, \, R_b^*= -{98 \over 219}, \, \, \,
R_{\tau}^* = -{285 \over 292}, \, \, \,
R_{\lambda}^* = 0, \, \, \, R_{\kappa}^* = -{ 1 \over 2},  \nonumber \\
R^{'}* &=& {164 \over 219}, \, \, \,
R^{''*} = {611 \over 876}, \, \, \,
\end{eqnarray}
which must also be rejected as being unphysical. {\it We conclude that $B$ and
$L$ violating couplings of the highest generation cannot simultaneously 
approach a non-trivial fixed point in the NMSSM}. This is one of the most 
important conclusions that we draw from the analysis of the 
renormalization group equations  of NMSSM, and is analogous to the 
corresponding result in MSSM~\cite{ap2}.

Given this result, it is now natural to consider the possibility of having 
either $B$ or $L$ violation, but not both simultaneously, involving
the trilinear couplings with highest generation indices in 
the RG evolution in the NMSSM.

\subsubsection{Fixed points with baryon number violation}
 
In this section we shall consider the infrared fixed points of the NMSSM
with $B$ violation. Thus, we shall consider the Yukawa couplings
$h_t, h_b, h_{\tau}, \lambda, \kappa$,  and the baryon number violating
coupling $\lambda^{''}_{233}$. We order the ratios of the trilinear couplings
to the gauge coupling $g_3$ 
as $R_i = (R_t, R_b, R_{\tau}, R_{\lambda}, R_{\kappa}, R^{''})$,
resulting in the anomalous dimension matrix\\

\begin{equation}\label{sb1}
S_{B1}=\left[
\begin{array}{c c c c c c}
6 & 1 & 0 & 1 & 0 & 2 \\
1 & 6 & 1 & 1 & 0 & 2 \\
0 & 3 & 4 & 1 & 0 & 0 \\
3 & 3 & 1 & 4 & 2 & 0 \\
0 & 0 & 0 & 6 & 6 & 0 \\
2 & 2 & 0 & 0 & 0 & 6 \\
\end{array}
\right],
\end{equation}
leading to the fixed point values
\begin{equation}
R_t^*={9\over 16}, \, \, \, R_b^*= {11 \over 16}, \, \, \,
R_{\tau}^* = -{5 \over 8}, \, \, \,
R_{\lambda}^* = -{41 \over 16}, \, \, \, R_{\kappa}^* = {33  \over 16},  \, \, \,
R^{''*} = {5 \over 12}, \, \, \,
\end{equation}
which must be rejected as being unphysical. We are, therefore, 
constrained to consider fixed point with one of the couplings approaching 
a zero fixed point value, with all others attaining a non-trivial
fixed point value. We first consider a fixed point with $R_{\tau}^* = 0$, 
with all others obtaining a non-zero fixed point value. In this case we obtain
the fixed point values
\begin{equation}
R_{\tau}^*=0, \, \, \, R_t^*= {5 \over 8}, \, \, \,
R_b^* = {5 \over 8}, \, \, \,
R_{\lambda}^* = -{23\over 8}, \, \, \, R_{\kappa}^* = {19  \over 8},  \, \, \,
R^{''*} = {5 \over 12}, \, \, \,
\end{equation}
which is a physically unacceptable fixed point solution. Proceeding in
this manner,
it can be shown that there is no acceptable infrared fixed point solution with
one of the couplings $h_t, h_b, h_{\tau}, \lambda, \kappa,  \lambda_{233}^{''}$
approaching a zero fixed point value, and 
the remaining approaching a non-zero fixed point value.

We are, therefore, led to the consideration of a fixed point in which two
of the Yukawa couplings approach a zero fixed point value, with all others 
attaining a non-zero fixed point value. We try the fixed point with
$R_{\lambda}^* = R_{\kappa}^* = 0$, with the rest approaching  
a non-zero fixed point value.  In this case we obtain the fixed point values
\begin{equation}
R_t^* = {31 \over 292}, \, \, \, R_b^* = {22 \over 73}, \, \, \, 
R_{\tau}^* = -{285\over 292}, \, \, \, R^{''*} = {611 \over 876},
\end{equation}
which must be rejected as a fixed point solution. Taking other pairs of 
Yukawa couplings to have a zero fixed point values, with the remaining 
obtaining non-zero values, we find that there is no acceptable fixed point
solution in this case.

We next consider the situation in which three of the couplings attain
zero fixed point values, with the remaining attaining a non-zero fixed
point value. Taking $R_{\tau}^* = R_{\lambda}^* = R_{\kappa}^* = 0$, we obtain
the fixed point
\begin{equation}
R_t^* = {2 \over 17}, \, \, \, R_b^* = {2 \over 17}, \, \, \, 
R^{''*} = {77 \over 102},
\label{baryon1}
\end{equation}
which is a physically acceptable fixed point solution. We find that there
is no other acceptable fixed point solution with three of the couplings
obtaining fixed point values, with the remaining approaching 
a non-trivial fixed point value.

We now consider the case when four of the trilinear Yukawa couplings 
approach a trivial fixed point, with the remaining two having a
non-trivial fixed point value. In this case we find the following
acceptable fixed point solutions:
\begin{equation}
R_{\tau}^* = R_{\lambda}^* = R_{\kappa}^* = R^{''*} = 0, \, \, \, 
R_t^* = {1\over 3}, \, \, \, R_b^* = {1 \over 3},
\label{baryon2}
\end{equation}
and
\begin{equation}
R_{\tau}^* = R_{\lambda}^* = R_{\kappa}^* = R_b^* = 0, \, \, \,
R_t^* = {1\over 8 }, \, \, \, R^{''*} = {19 \over 24},
\label{baryon3}
\end{equation}                                     

Since there are more than one theoretically acceptable IRFPs in this
case, it is necessary to determine, which, if any, of these fixed points
is more likely to be realized in nature.  To this end, we must examine
the stability of each of the fixed point solutions (\ref{baryon1}),
(\ref{baryon2}) and (\ref{baryon3}).

The infra-red stability of a fixed point is determined by the sign of
the quantitites
\begin{equation}\label{lambdaequation}
\lambda_i={1\over b_3}\left[\sum_{j=m+1}^n S_{ij} R^*_j - (r_i+b_3)\right],
\end{equation}
for those couplings which have  a fixed point value  zero, 
$R_i^*=0,\, i = 1,2,...,m,$ and by the sign of the eigenvalues of the
matrix
\begin{equation}\label{aijmatrix}
A_{ij}={1\over b_3} R^*_i S_{ij}, \, i=m+1,...,n,
\end{equation}
for those couplings  which have a non-trivial IRFP~\cite{allanach},
where $R_i^*$ is the set of the non-trivial 
fixed point values of the
Yukawa couplings under consideration, and $S_{ij}$ is the matrix
appearing in the corresponding renormalization group
equations (\ref{RI})
for the ratios $R_i$.  For stability, we require all the $\lambda_i$'s
to have negative sign,
and the eigenvalues
of the matrix (\ref{aijmatrix}) to have negative real parts
(note that the QCD $\beta$-function $b_3$ is negative).
For the infra-red fixed point (\ref{baryon1}), 
we find from Eq.~(\ref{lambdaequation})
\begin{equation}\label{baryon1stability1}
\lambda_1=-{19\over 17}, \, \, \, \lambda_2 = -{21\over 17},
\, \, \, \lambda_3 = -1,
\end{equation}
corresponding to $R^*_\tau=0, R_{\lambda}^* = 0,$ and 
$R_{\kappa}^* = 0$, 
respectively and for the eigenvalues of 
the matrix (\ref{aijmatrix})
\begin{equation}\label{baryon1stability2}
\lambda_4=-{10\over 51} \, \, \simeq -0.2, \, \, \,
\lambda_5={-273 - \sqrt{43113}\over 306} \, \, \simeq -1.6, \, \, \,
\lambda_6={-273+\sqrt{43113}\over 306} \, \, \simeq -0.2,
\end{equation}
corresponding to the non-trivial fixed point values for $R''^*,\,
R^*_b$ and $R^*_t,$ respectively.  Since all the $\lambda_i,\, i=1,2,3,4,
5,6$ are negative, the fixed point (\ref{baryon1}) is infra-red stable.

Next we consider the stability of the fixed point (\ref{baryon2}).
Since in this case $R_{\lambda}^* = R_{\kappa}^* = R_\tau^*=R''^*=0$, 
we have to obtain the behaviour
of these  couplings around the origin.  This behaviour is determined by the
quantities (\ref{lambdaequation}) which, in this case,  are
\begin{equation}
\lambda_1=-{4\over 3}, \, \lambda_2=-{5\over 3}, \, \lambda_3 = -1\,
\lambda_4 = {11\over 9},
\end{equation}
thereby indicating that the fixed point is unstable in the infra-red region.
For completeness we also obtain the behaviour of $R_b$ and $R_t$ around
their respective fixed points governed by the corresponding
eigenvalues of the matrix (\ref{aijmatrix}). We obtain for
the eigenvalues of this matrix
\begin{equation}
\lambda_5 =-{5\over 9}, \, \lambda_6=-{7\over 9}.
\end{equation}
Although $\lambda_5,\, \lambda_6$ are negative, the fact 
that $\lambda_4>0$ implies that the fixed point
(\ref{baryon2}) is unstable in the infra-red region.
Thus, this infra-red fixed point with trivial
fixed point value for the baryon number
violating coupling $\lambda_{233}^"$ will 
never be realized at low energies, 
and must be rejected.  
Similarly,  it is straightforward to see that the fixed point
(\ref{baryon3}) is  unstable in the infra-red, and must be rejected.

One may also consider the case where the couplings $\lambda''^*_{233}, \,
\kappa, \, \lambda, \, h_\tau,\, h_b$ attain trivial fixed point values, 
whereas $h_t$ attains
a non-trivial fixed point value.  In this case we find $R_t^*=7/18$,
the well-known Pendleton-Ross type 
fixed point~\cite{pendross}.   The stability
of this fixed point solution is obtained by simply considering
the quantities (\ref{lambdaequation})
\begin{equation}
\lambda_i={1\over b_3} \left[ (S_{B1})_{i6} R_6^* -(r_i+b_3)\right],
\, \, \, \,
i=1,2,3,4,5,
\end{equation}
where the matrix $S_{B1}$ is the one which appears in (\ref{sb1}) but with
the ordering of the ratios as $R_i = (R_{\tau}, R_{\lambda}, R_{\kappa},
R_b, R^{''}, R_t)$,
which yields
\begin{equation}
\lambda_1=-1, \, \lambda_2 = -{25\over 18}, \,
\lambda_3 = -1, \lambda_3={35\over 54}, \, 
\lambda_5={38\over 27}, 
\end{equation}
thereby rendering this fixed point unstable.

Finally, one may consider the case where $R''^*=0$, with $R_t, \, R_b, \,
R_\tau,\, R_{\lambda}, R_{\kappa}$,
attaining non-trivial fixed point values.  This is the case
of the NMSSM with $R_p$ conservation and 
with all the third generation Yukawa couplings taken into
account.  In this case, we find the fixed point solution:
\begin{equation}
R^*_t={37\over 48},  R^*_b={33\over 48} R_\tau^*=-{5 \over 8}, \, 
R^*_{\lambda}=-{51\over 16}, \, R^*_k={43\over 16},
\end{equation}
which must be rejected as being unphysical.

Thus, we have shown that the only fixed point
which is stable in the infra red region is the
baryon number,  and $R_p$,  violating solution (\ref{baryon1}).  
We note that the value
of $R_t^*$ corresponding to this solution is lower than the 
fixed point value of $7/18$ for NMSSM (and MSSM) with baryon number,
and $R_p$,  conservation. 

\bigskip

\subsubsection{Fixed points with lepton number violation}

In order to investigate the possibility of having a stable fixed point
with lepton number violation, we now  study  the renormalization
group evolution for the lepton number, and $R_p$, violating couplings
in the superpotential (\ref{Lviolating}). We shall consider the 
trilinear couplings $\tilde\lambda_3, \lambda_{233}$ and $\lambda_{333}^{'}$,
together with the Yukawa couplings $h_t, h_b, h_{\tau}, \lambda, \kappa$.
As indicated in   Sec. II, we shall assume, 
without loss of generality, that there is a hierarchy of lepton
number violating couplings, and consider only one lepton number violating
coupling at
a time. Thus, we shall consider three different cases, i.e., we shall take
either $\tilde\lambda_3 \gg \lambda_{233}, \lambda'_{333}$,
or $\lambda_{233} \gg \tilde\lambda_3, \lambda'_{333}$,
or $\lambda'_{333} \gg \tilde\lambda_3, \lambda_{233}$, 
respectively.

We first consider the case when $\tilde\lambda_3$ is the
dominant lepton number violating coupling. Ordering the
ratios of the squares of the trilinear couplings to the
square of the gauge coupling $g_3$ as 
$R_i = (R_t, R_b, R_{\tau}, R_{\lambda}, R_{\kappa}, \tilde R_3)$,
the anomalous dimension matrix, $S_{L11}$, in this case
is given by
\begin{equation}
S_{L11}=\left[
\begin{array}{c c c c c c}
6 & 1 & 0 & 1 & 0 & 1\\
1 & 6 & 1 & 1 & 0 & 0\\
0 & 3 & 4 & 1 & 0 & 1\\
3 & 3 & 1 & 4 & 2 & 4\\
0 & 0 & 0 & 6 & 6 & 6\\
3 & 0 & 1 & 4 & 2 & 4\\
\end{array}
\right],
\label{lepton11}
\end{equation}
leading to the fixed point values
\begin{equation}
R^*_t={32 \over 45}, \, R_b^* = 0, \,  R^*_{\tau} = -{4\over 15 }, \, 
R_{\lambda}^* = {17\over 9}, \, R_{\kappa}^* = {43 \over 30},\,
\tilde R_3^* = -{172\over 45}.
\end{equation}
Since $R^*_\tau, \tilde R^*_3 < 0$, this fixed point is unacceptable.
Thus, there is no nontrivial 
infrared fixed point for both the
lepton number violating 
coupling $\tilde\lambda_3$, and all other trilinear couplings in the NMSSM.

Next we consider we consider a trivial fixed point for one of the
couplings, and nontrivial one for the. Taking
$R_b^* = 0$, which is relevant for low values of $\tan\beta$,
the anomalous dimension matrix is
\begin{equation}
S_{L12}=\left[
\begin{array}{c c c c c}
6 &  0 & 1 & 0 & 1\\
0 &  4 & 1 & 0 & 1\\
3 &  1 & 4 & 2 & 4\\
0 &  0 & 6 & 6 & 6\\
3 &  1 & 4 & 2 & 4\\
\end{array}
\right],
\label{lepton12}
\end{equation}
which is singular. This solution corresponds to a fixed line or a surface
\footnote{Here we correct an error in \cite{paulraj}}.
Continuing in this manner, and taking each one of the couplings to have
a trivial fixed point value with the rest a nontrivial fixed point value,
we find there is no acceptable infrared fixed point in this case.

We now try to obtain a fixed point with two of the couplings approaching 
a trivial fixed point, while the rest having a non-trivial fixed point. 
Taking  $R_{\lambda}^* = R_{\kappa}^* = 0$, we obtain the fixed point values
for the remaining couplings
\begin{equation}
R^*_t={11\over 24}, \, \, \, R^*_b={11\over 24}, \, \, \,
R^*{\tau}=-{7\over 8}, \, \, \, \tilde R^*_3=-{7\over 8}, 
\end{equation}
which is an unacceptable fixed point. Continuing  in this manner
we find that there are no acceptable infrared fixed points in this case.
We have also checked that for the case with three or four
of the couplings 
approaching a zero fixed point value, with the remaining having
non-trivial fixed point values, there are no physically
acceptable fixed point solutions. Finally, for the case with all the couplings,
except the top-quark Yukawa coupling, approaching a trivial fixed point value,
we have
\begin{equation}
R_b^* = R_{\tau}^* = R_{\lambda}^* = R_{\kappa}^* 
= \tilde R_3^* = 0,\\ \nonumber
R_t^* = {7\over 18},
\end{equation}
which is an acceptable fixed point. However, it is easily seen that this 
fixed point is unstable in the infrared region. We, therefore, conclude 
that for the case with  lepton number violation through the coupling
$\tilde \lambda_3$, there are no acceptable infra-red fixed points.

If on the other hand the coupling $\lambda_{233}$
is the dominant of the lepton number couplings, then 
the anomalous dimension matrix is given by
\begin{equation}
S_{L_{21}}=\left[
\begin{array}{c c c c c c}
6 & 1 & 0 & 1 & 0 & 0\\
1 & 6 & 1 & 1 & 0 & 0\\
0 & 3 & 4 & 1 & 0 & 4\\
3 & 3 & 1 & 4 & 2 & 0\\
0 & 0 & 0 & 6 & 6 & 0\\
0 & 0 & 4 & 0 & 0 & 4\\
\end{array}
\right],
\label{lepton21}
\end{equation}
in the basis $R_i = (R_t, R_b, R_{\tau}, R_{\lambda}, R_{\kappa}, R)$,
leading to the fixed point values
\begin{equation}
R_t^* = {17 \over 24}, \, \, R_b^* = {23 \over 24}, \, \, 
R_{\tau}^* = -{5 \over 4}, \, \, 
R_{\lambda} = -{23\over 8}, \, \, R_{\kappa}^* = {19 \over 8} \, \, 
R^* = {1 \over 2},
\end{equation}
which must be rejected as a fixed point. We can try with one of the couplings,
namely $R_b^*$ approaching a trivial fixed point value with the rest 
approaching a nontrivial fixed point. This case is relevant for low values of
$\tan\beta$. We get
\begin{equation}
R_b^* = 0, \, \, R_t^* = {7 \over 18 }, \, \,  
R_{\tau}^* = -{ 19\over 6}, \, \, 
R_{\lambda} = 0, \, \, R_{\kappa}^* = -{1 \over 2} \, \, 
R^* = {29 \over 12},
\end{equation}
which is also unacceptable. We may also try a fixed point with 
$R_{\tau}^* = 0$, and a non-trivial fixed point for other couplings. 
In this case we get
\begin{equation}
R_{\tau}^* = 0, \, \, R_t^* = { 5\over 6 }, \, \, R_b^* = {5 \over 6}, \, \,
R_{\lambda} = -{7 \over 2}, \, \, R_{\kappa}^* = 3 \, \,
R^* = -{3 \over 4},
\end{equation}
making this also an unacceptable fixed point. Continuing in this fashion,
we find that there are no fixed points with one of the couplings approaching
a zero fixed point value, with the rest having non-trivial fixed point values.

We now try to obtain fixed points with two of the couplings having a
fixed point value zero, and the rest having non-zero fixed point values.
Taking $R_b^* = R_{\tau}^* = 0$, we obtain
\begin{equation}
R_t^* = {20 \over 27 }, \, \,
R_{\lambda} = -{19 \over 9}, \, \, R_{\kappa}^* = {29 \over 18} \, \,
R^* = -{3 \over 4},
\end{equation}
which is unphysical, and hence unacceptable. Similarly, we have checked 
there is no physically acceptable fixed point with any of the
two couplings approaching a trivial fixed point value, 
with the rest attaining a non-trivial fixed point value. Continuing 
in this fashion, we find that there are no acceptable fixed points
in the infrared region for the lepton number violating coupling 
$\lambda_{233}$. In particular, the fixed point with five of the couplings
approaching a trivial fixed point value
\begin{equation}
R_b^* = R_{\tau}^* =  R_{\lambda} = R_{\kappa}^* = R^* = 0, 
\end{equation}
and $R_t$ approaching a non-trivial fixed point value, 
$R_t^* = 7/18$, is unstable in the infrared region.

Finally, we consider the case when the
coupling $\lambda'_{333}$ is the dominant of the lepton number violating
couplings. With the ordering of the couplings $R_i = (R_t, R_b, R_{\tau},
R_{\lambda}, R_{\kappa}, R')$, the anomalous dimension matrix in this case is
\begin{equation}
S_{L31}=\left[
\begin{array}{c c c c c c}
6 & 1 & 0 & 1 & 0 & 1\\
1 & 6 & 1 & 1 & 0 & 6\\
0 & 3 & 4 & 1 & 0 & 3\\
3 & 3 & 1 & 4 & 2 & 0\\
0 & 0 & 0 & 6 & 6 & 0\\
1 & 6 & 1 & 0 & 0 & 0\\
\end{array}
\right],
\label{lepton31}
\end{equation}
leading to the fixed point values
\begin{equation}
R_t^* = {55 \over 183}, R_b^* = -{107\over 183}, 
R_{\tau}^* = -{70 \over 61},  R_{\lambda} = 0, R_{\kappa}^* = -{1\over 2},
R'^* = {68 \over 61},
\end{equation}
which are unacceptable. As in the other cases, we try a fixed point with one of the couplings approaching a zero fixed point value, and the rest non-trivial 
values. Taking, $R_b^* = 0$, we have
\begin{equation}
R_t^* = {121 \over 210}, 
R_{\tau}^* = -{47 \over 70},  R_{\lambda} = -{107\over 70},
R_{\kappa}^* = {36\over 35},
R'^* = {17 \over 42},
\end{equation}
which is an unacceptable fixed point. Continuing in this manner, we find that 
in this case there are only two acceptable infrared fixed points:
\begin{equation}
R_b^* =  R_{\tau}^* = R_{\lambda} = R_{\kappa}^* = 0, \, \, \, \, \, \, 
R_t^* = {1 \over 3}, \, \, R'^* = {1 \over 3}, \label{lprimea}
\end{equation}
and
\begin{equation}
R_{\tau}^* = R_{\lambda} = R_{\kappa}^* = R' = 0, \, \, \, \, \, \, 
R_t^* = {1 \over 3}, \, \, R_b^* = {1 \over 3}. \label{lprimeb}
\end{equation}
In order to determine which of these fixed points, if any, is actually
realised, we need to check the stability of these fixed points. We first 
consider the fixed point (\ref{lprimea}). In this case the quantities 
(\ref{lambdaequation}) are calculated to be
\begin{equation}
\lambda_1 = 0, \, \, \lambda_2 = -{4 \over 3}, \, \,
\lambda_3 = -{4 \over 3}, \, \, \lambda_4 = -1
\label{stability1}
\end{equation}
corresponding to the the zero fixed point values in (\ref{lprimea}).
From (\ref{stability1}) we conclude that the fixed point (\ref{lprimea})
will never be reached in the infrared region. The fixed point is either
a saddle point or an ultra-violet istable fixed point. 

The stability of the fixed point (\ref{lprimeb}) is determined in a
manner analogous to that of the fixed point (\ref{lprimea}). We find
that the fixed point (\ref{lprimeb}) is either a saddle point, or
an ultra-violet stable fixed point. 
That the 
stability properties of the fixed points
(\ref{lprimea}) and (\ref{lprimeb})
are identical is a consequence of the symmetry of
the renormalization group 
equations ~(\ref{HT}) - (\ref{DLD}).
We conclude that there are no non-trivial stable fixed points in the 
infra-red region for the $L$ violating coupling $\lambda'_{333}$.

To sum up, we have found that there are no IRSFPs in the NMSSM with
the highest generation lepton number  violation.  
This result, together with the result on the fixed point with 
baryon number violation, shows that
only the simultaneous non-trivial fixed point (\ref{baryon1})
for the baryon number violating coupling
$\lambda''_{233}$, and the top- and bottom-quark Yukawa couplings,
$h_t$ and $h_b$,  is stable in the infra-red region.
This result is analogous to the result obtained in MSSM, and shows that
the additional couplings in NMSSM have no effect on the infrared
fixed point behavior of the Yukawa and baryon number violating couplings.

It is appropriate to examine the implications of the value  of  the
top-quark mass predicted by our fixed point analysis. From (\ref{baryon1}),
it is readily seen that the fixed point value for the
top-quark Yukawa coupling translates into a top-quark (pole) mass
of about $m_t \simeq 70\sin\beta$ GeV, which is incompatible with the
measured value of~\cite{pdg} of top mass, $m_t \simeq 174$ GeV, for any 
value of  $\tan\beta$. It follows that the true fixed point
obtained here provides only a lower bound on the baryon number
violating coupling $\lambda_{233}'' \stackrel{>}{\sim} 0.97$.

\subsection{Infrared fixed points for the trilinear soft supersymmetry
breaking parameters}

Having obtained the infrared fixed point structure of the Yukawa 
couplings of  NMSSM
with baryon and lepton number violation, we now consider the 
renormalization group evolution and the fixed point
structure for the 
soft supersymmetry breaking trilinear parameters
$A_i$.  Since there is only one IRSFP in
this case, we shall consider the
IRFPs for the $A$ parameters corresponding to this case only, i.e.
for $A_t$, $A_b$, $A_\tau$, $A_{\lambda}, A_{\kappa}$,  
and $A_{\lambda''_{233}}$.

Retaining only these parameters, and with the 
definitions $\tilde{A}_i=A_i/M_3$
$(A_i=A_t,  A_b,  A_\tau,  A_{\lambda},  A_{\kappa},  A_{\lambda''_{233}})$, 
we obtain from
Eq.~(\ref{rge7}) - Eq.~(\ref{rge15}) the renormalization group
equations for the relevant
$\tilde{A}_i$ (neglecting the $SU(2)_L$ and $U(1)_Y$ 
gauge couplings):
\begin{eqnarray}
{d \tilde{A}_t\over d(-\ln \mu^2)} &=&
\tilde{\alpha}_3 \left[ {16\over 3} - (6 R_t -b_3) \tilde{A}_t-
R_b \tilde{A}_b -R'' \tilde{A}_{\lambda''_{233}} \right]  \label{atildet}, \\
{d \tilde{A}_b\over d(-\ln \mu^2)} &=&
\tilde{\alpha}_3 \left[ {16\over 3} - R_t  \tilde{A}_t-
(6 R_b- b_3) \tilde{A}_b - R_\tau \tilde{A}_\tau 
- R_{\lambda}\tilde A_{\lambda} 
- 2R'' \tilde{A}_{\lambda''_{233}} \right]  \label{atildeb}, \\
{d \tilde{A}_\tau\over d(-\ln \mu^2)} &=&
\tilde{\alpha}_3 \left[ 
-3 R_b \tilde{A}_b - (4 R_\tau -b_3) \tilde{A}_\tau - 
 R_{\lambda} \tilde A_{\lambda} \right],  
\label{atildetau} \\
{d \tilde{A}_\lambda\over d(-\ln \mu^2)} &=&
\tilde{\alpha}_3 \left[
-3 R_t \tilde A_t - 3 R_b \tilde A_b - R_{\tau} \tilde A_{\tau}
- (4 R_{\lambda} - b_3) \tilde A_{\lambda} - 2 R_{\kappa} \tilde A_{\kappa}
\right ], \label{atildelambda} \\
{d \tilde{A}_{\kappa}\over d(-\ln \mu^2)} &=&
\tilde{\alpha}_3 \left[
-6 R_{\lambda} \tilde A_{\lambda} - (6 R_{\kappa} - b_3) \tilde A_{\kappa}
\right ], \label{atildek}\\
{d \tilde{A}_{\lambda''_{233}}\over d(-\ln \mu^2)} &=&
\tilde{\alpha}_3 \left[ 8 - 2 R_t  \tilde{A}_t- 2 R_b \tilde{A}_b 
-(6 R''-b_3) \tilde{A}_{\lambda''_{233}} \right],  \label{atildelambdadpr} 
\end{eqnarray}
which can be written in a compact form 
\begin{equation}\label{acomponent}
{d \tilde A_i \over d(-\ln \mu^2)}=\tilde{\alpha}_3\left[r_i
-\sum_j K_{ij} \tilde{A}_j \right],
\end{equation}
in the basis
$\tilde A_i = (\tilde A_{\tau}, \tilde A_{\lambda}, \tilde A_{\kappa},
\tilde A_{\lambda''_{233}}, \tilde A_b, \tilde A_t)$, 
where $r_i = (0, 0, 0, 8, 16/3, 16/3)$,
and where $K$ is a matrix whose
entries are fully specified by the wave function anomalous dimensions
and $R_i$.  A fixed point for $\tilde A_i$ is, then,
reached when the right hand side
of Eq.~(\ref{acomponent}) vanishes for all $i$.  Denoting this fixed
point solution by $\tilde{A}^*_i$, we have
\begin{equation}\label{afixedpoint}
r_i-\sum_j K^*_{ij} \tilde{A}_j^*=0,
\end{equation} 
where $K^*$ is the matrix $K$ evaluated when $R_i$ take their
fixed point values $R^*_i$.  
From Eq.~(\ref{atildet}) - Eq.~(\ref{atildelambdadpr}) we see that
the matrix $K$ at the fixed point  can be written as 
\begin{equation}
K^*=\left[
\begin{array}{c c c c c c}
4 R_\tau^*-b_3 & R_{\lambda}^* & 0 & 0 & 3 R_b^* & 0 \\
R_{\tau}^* & (4 R_{\lambda}^* - b_3) & 2 R_{\kappa}^* & 0 & 3 R_b^* & 3 R_t^*\\
0 & 6 R_{\lambda}^* & (6 R_{\kappa}^* - b_3) & 0 & 0 & 0\\
0 & 0 & 0 & (6 R''^*-b_3) & 2 R_b^* & 2 R_t^*  \\
R_\tau^* & R_{\lambda}^* & 0 & 2 R''^* & (6 R_b^* -b_3) & R_t^* \\
0 & 0 & 0 & 2 R''^* & R^*_b & (6 R_t^* - b_3) \\
\end{array}
\right],
\end{equation}
with $R_\tau^* = R_{\lambda}^* =  R_{\kappa}^* = 0$, and
$R''^*, R^*_b, R^*_t$ given by their fixed point
values in (\ref{baryon1}).  The fixed points $\tilde{A}_i^*$ are given by
the solution of 
\begin{equation}\label{asolution}
\tilde{A}_i^*=\sum_j (K^{*-1})_{ij} r_j,
\end{equation}
with the result
\begin{equation} \label{truetrilinear}
\tilde{A}_\tau^*=-{2\over 17}, \, \, \tilde A_{\lambda}^* = -{4 \over 17},
\, \, \tilde A_{\kappa}^* = 0, \, \, \tilde{A}_{\lambda''_{233}}^*
=\tilde{A}_b^*=\tilde{A}_t^*=1.
\end{equation}
We have carried out the stability analysis of the 
fixed point
(\ref{truetrilinear}) in a manner analogous to that of the Yukawa couplings
in the previous subsection, and find it to be infra-red stable.
We note that the fixed point values of $\tilde A_{\tau}, 
\tilde A_{\lambda''_{233}}, \tilde A_b,
\tilde A_t$ in (\ref{truetrilinear}) are the same as in MSSM with
baryon number violation~\cite{ap2}.

\section{Quasi-Fixed Points}

The infrared fixed points 
that we have discussed in the previous section
are the true IRFPs of the renormalization group equations.
However, these fixed points are not reached in practice, the
range between the large(GUT) scale and the weak scale being too small
for the ratios to closely approach the fixed point values.  In that case,
the various couplings may be determined by the quasi-fixed point 
behaviour~\cite{hill}, where the value of various couplings at
the weak scale is independent of its value at the large(GUT)
scale, provided the couplings at the unification scale are large.  
In this section, we shall discuss the quasi-fixed point behaviour of the
Yukawa couplings and the $A$ parameters of the NMSSM with $B$ 
violation, corresponding to the true fixed points 
that we have obtained in the previous section.

In order to discuss the quasi-fixed points, we shall first 
write down the analytical solution for the trilinear
couplings of the NMSSM in a closed form.
Since the simultaneous fixed point for the third generation 
Yukawa couplings, $\lambda$, $\kappa$, and the baryon number violating
coupling $\lambda''_{233}$ is stable, we shall consider  the
quasi-fixed points for these couplings only.
For this purpose we define
\begin{equation}
\tilde Y_t =  {h_t^2\over {16{\pi}^2}}, \, \, \,  
\tilde Y_b =  {h_b^2\over {16{\pi}^2}}, \, \, \,
\tilde Y_{\tau} =  {h_{\tau}^2\over {16{\pi}^2}},
\label{def1}
\end{equation}
\begin{equation} 
\tilde Y_{\lambda} = {\lambda^2\over {16{\pi}^2}},  \hspace{1cm}
\tilde Y_{\kappa} = {\kappa^2\over {16{\pi}^2}}, \label{def2}
\end{equation}
\begin{equation}
\tilde Y'' = {\lambda^{''2}_{233}\over {16{\pi}^2}}.
\label{def3}
\end{equation}
Then the solution of the RG equations (\ref{HT}) - 
(\ref{DKA}), and (\ref{DLD})
for the Yukawa, the trilinear couplings $\lambda$ and $\kappa$,
and the $B$  violating
couplings can be written in a closed form~\cite{auberson1}
\begin{equation}
\tilde Y_i(t) = { {\tilde Y_i(0) F_i(t)}\over{1 + a_{ii}
\tilde Y_i(0)\int_0^t F_i(t') dt'}}, \, \, \, \, \, \, \,
t = ln({{M_G^2}\over{\mu^2}}), \label{yukawasolution}
\end{equation}
where $M_G$ is some large initial scale, and 
where $\tilde Y_i$ stands for the functions $\tilde Y_t~, \,
\tilde Y_b~, \,  \tilde Y_{\tau}, \, \tilde Y_{\lambda}, \,
\tilde Y_{\kappa}$, and  $\tilde Y''$, respectively.
Analogous notation holds for the functions $F_i$.
The quantities $a_{ii}$ are the 
diagonal elements of the wave function anamolous dimension matrix, 
and are given by
\begin{equation}
a_{ii} = \{6, \, 6, \, 4, \, 4, \, 6, \,  6\}, \label{aquantities}
\end{equation} 
and the functions $F_i$ are given by the set of integral equations
\begin{small}
\begin{eqnarray}
F_t(t) &=& {{E_t(t)}\over{(1 + 6 \tilde Y_b(0)\int_0^t F_b(t') dt')^{1/6} 
(1 + 4 \tilde Y_{\lambda}(0)\int_0^t F_{\lambda}(t') dt')^{1/4} 
(1 + 6 \tilde Y''(0)\int_0^t F''(t') dt')^{1/3}}}, \label{ftequation}\\
F_b(t) &=& {{E_b(t)}\over{(1 + 6 \tilde Y_t(0)\int_0^t F_t(t') dt')^{1/6} 
(1 + 4 \tilde Y_{\tau}(0)\int_0^t F_{\tau}(t') dt')^{1/4}
(1 + 4  \tilde Y_{\lambda}(0)\int_0^t F_{\lambda}(t') dt')^{1/4}}}\nonumber \\ 
& \times &
{{1} \over {(1 + 6 \tilde Y''(0)\int_0^t F''(t') dt')^{1/3}}}, 
\label{fbequation}\\
F_{\tau}(t) &=& {{E_{\tau(t)}}\over
{(1 + 6 \tilde Y_b(0)\int_0^t F_b(t') dt')^{1/2} 
(1 + 4 \tilde Y_{\lambda}(0)\int_0^t F_{\lambda}(t') dt'){1/4} }}, 
\label{ftauequation}\\
F_{\lambda}(t) &=& {{E_{\lambda}(t)}\over
{(1 + 6 \tilde Y_t(0)\int_0^t F_t(t') dt')^{1/2}
(1 + 6 \tilde Y_b(0)\int_0^t F_b(t') dt')^{1/2}
(1 + 4 \tilde Y_{\tau}(0)\int_0^t F_{\tau}(t') dt')^{1/4}}}\nonumber \\
& \times & 
{{1}\over
{(1 + 6 \tilde Y_{\kappa}(0)\int_0^t F_{\kappa}(t') dt')^{1/3}}}, 
\label{flambdaequation}\\
F_{\kappa}(t) &=& {{E_{\kappa}(t)}\over
{(1 + 4 \tilde Y_{\lambda}(0)\int_0^t F_{\lambda}(t') dt')^{3/2}}},
\label{fkequation} \\
F''(t) &=& {{E''(t)}\over
{(1 + 6 \tilde Y_t(0)\int_0^t F_t(t') dt')^{1/3}
(1 + 6 \tilde Y_b(0)\int_0^t F_t(t') dt')^{1/3}}},
\label{fdoubleprimeequation}
\end{eqnarray}
\end{small}
where the functions $E_i(t)$ 
$( = E_t(t), E_b(t), E_{\tau}(t), E_{\lambda}(t),  E_{\kappa}(t)$ and
$E''(t) )$ are given by
\begin{equation}
E_j(t) = \prod_{m =1}^3 \left( 1 + b_m \tilde \alpha_m(0) t\right)^{c_{jm}/b_m},
\label{eequation}
\end{equation}
with
\begin{equation}
\tilde\alpha_m(0) = {{g_m^2(0)}\over{16 \pi^2}}, \, \, \, m  = 1, \, 2, \, 3, 
\label{tildealpha}
\end{equation}
\begin{equation}
c_{tm} = \left( {{13}\over {15}}, \, 3, \, {{16}\over {3}} \right), \, \, \, \,
c_{bm} = \left( {{7}\over {15}}, \, 3, \, {{16}\over {3}} \right), \, \, \, \,
c_{{\tau}m } = \left( {{9}\over{ 5}}, \, 3, \, 0 \right), \, \, \, \, \,
\label{cfunctions1}
\end{equation}
\begin{equation} 
c_{{\lambda}i} = \left( {{3}\over{5}}, \, 3, \, 0 \right), 
\, \, \, \, \, \,
c_{{\kappa} i} = \left( 0, \, 0, \, 0 \right), 
\, \, \, \, \,
c_{{\lambda''_{233}}i} = \left( {{4}\over{5}}, \, 0, \, 8 \right).
\label{cfunctions2}
\end{equation}
The solutions for the RG equations for the gauge couplings 
(\ref{gauge1}) and the 
gaugino masses (\ref{gaugino1}) are well known and will not be repeated here.
We note that (\ref{yukawasolution}) gives the exact solution for the Yukawa couplings,
while $F_i$'s in (\ref{ftequation}) - (\ref{fdoubleprimeequation}) should 
in principle be solved iteratively.

In the regime where the Yukawa couplings
$\tilde Y_t(0), \, \, \tilde Y_b(0), \, \, \tilde Y_{\tau}(0), \, \,
\tilde Y_{\lambda}(0), \, \, \tilde Y_{\kappa}(0), \, \, \tilde Y''(0) \, \, 
\rightarrow \infty$ with their ratios  fixed, it is legitimate to drop
$1$ in the denominators of the equations
(\ref{yukawasolution}) and (\ref{ftequation}) -- (\ref{fdoubleprimeequation}),
so that the exact solutions for the trilinear  couplings approach the 
infrared quasi-fixed-point (IRQFP) defined by
\begin{equation}
\tilde Y_i^{QFP}(t) = {{F_i^{QFP}(t)}\over{a_{ii}\int_0^t F_i^{QFP}(t') dt'}}, 
\label{quasiyukawa}
\end{equation}
with
\begin{eqnarray}
F_t^{QFP}(t) &=& {{E_t(t)}\over{(\int_0^t F_b^{QFP}(t') dt')^{1/6}
(\int_0^t F_{\lambda}^{QFP}(t') dt')^{1/4} (\int_0^t F''^{QFP}(t') dt')^{1/3}}},
\label{qftequation}\\
F_b^{QFP}(t) &=& {{E_b(t)}\over{(\int_0^t F_t^{QFP}(t') dt')^{1/6}
(\int_0^t F_{\tau}^{QFP}(t') dt')^{1/4}
(\int_0^t F_{\lambda}^{QFP}(t') dt')^{1/4}
(\int_0^t F''^{QFP}(t') dt')^{1/3}}},
\label{qfbequation}\\
F_{\tau}^{QFP}(t) &=& {{E_{\tau(t)}}\over
{(\int_0^t F_b^{QFP}(t') dt')^{1/2}
(\int_0^t F_{\lambda}^{QFP}(t') dt'){1/4}}},
\label{qtauequation}\\
F_{\lambda}^{QFP}(t) &=& {{E_{\lambda}(t)}\over
{(\int_0^t F_t^{QFP}(t') dt')^{1/2}
(\int_0^t F_b(t')^{QFP} dt')^{1/2}
(\int_0^t F_{\tau}(t')^{QFP} dt')^{1/4}
(\int_0^t F_{\kappa}^{QFP}(t') dt')^{1/3}}}, \label{qlambdaequation}\\
F_{\kappa}^{QFP}(t) &=& {{E_{\kappa}(t)}\over
{(\int_0^t F_{\lambda}^{QFP}(t') dt')^{3/2}}},
\label{qkequation} \\
F''^{QFP}(t) &=& {{E''(t)}\over
{(\int_0^t F_t^{QFP}(t') dt')^{1/3}
(\int_0^t F_b^{QFP}(t') dt')^{1/3} }}.
\label{qfdoubleprimeequation}
\end{eqnarray}
We stress here that both the dependence on the initial conditions
for each Yukawa coupling as well as the dependence on the 
ratios of initial values of Yukawa couplings  have completely dropped
out of the runnings in Eqs.(\ref{quasiyukawa}) and (\ref{qftequation})
-- (\ref{qfdoubleprimeequation}). In other words, the 
quasi-fixed-points (\ref{quasiyukawa})  are independent of
whether the $B$ and $L$ violating couplings and the third generation
Yukawa couplings are unified or not. The fact that the ratios of the various
Yukawa couplings do not enter 
Eqs. (\ref{quasiyukawa}) -- (\ref{qfdoubleprimeequation}) implies that 
these results are valid for any $\tan\beta$ regime.

One can also obtain the complete analytic solutions of the RG equations
for the trilinear supersymmetry breaking parameters $A_i$ in an analogous
manner. The  expressions for these are lengthy, and will not be written here.
Since the solutions for the quasi-fixed points must be obtained
either iteratively from (\ref{quasiyukawa}), or numerically from the
RG equations, we shall instead study the numerical solutions
for the quasi-fixed points for these $A$ parameters, as well as 
the Yukawa couplings,  in the following.

\subsection{Quasi-fixed points for the Yukawa couplings}

In order to determine the quasi-fixed points, one can solve
the RG equations for the various couplings numerically. We shall
do so in the next section. However, before doing so it is instructive
and useful to
obtain an analytical estimate of the quasi-fixed point values for these 
couplings. As stated above, we shall study the
quasi-fixed point 
for the couplings $h_t, h_b, h_{\tau}, \lambda,
\kappa, \lambda''_{233}$, since only the fixed point
(\ref{baryon1}) corresponding to these couplings is stable.

Writing the RG equations for (\ref{HT}) - (\ref{DKA}), and (\ref{DLP})
for these couplings in terms of $\tilde Y's$ defined 
in (\ref{def1}) - (\ref{def3}),
the existence of quasi-fixed point requires~\cite{barger}
\begin{equation}\label{approximatecondition}
{d \tilde Y_t \over dt}\simeq{d \tilde Y_b \over dt}
\simeq{d \tilde Y_\tau \over dt}\simeq{d \tilde Y_{\lambda} \over dt}
\simeq {d \tilde Y_{\kappa} \over dt} \simeq {d \tilde Y'' \over dt}
\simeq 0.
\end{equation}
The RG equations for $Y_{\lambda}$ and $\tilde Y_{\kappa}$  lead to
\begin{equation}
\tilde Y_{\lambda} = \tilde Y_{\kappa} = 0,
\end{equation}
thereby implying that the quasi-fixed point values for $\lambda$ and
$\kappa$ are zero. Substituting these values in the remaining RG equations, 
we get
\begin{equation}
\tilde Y_\tau^*={3\over 730}\left[121 \tilde \alpha_1 
+ 100 (\tilde \alpha_2-\tilde \alpha_3)\right]
\simeq -0.0016, 
\end{equation}
where we have used  $\alpha_1\simeq 0.017,\, \alpha_2\simeq
0.033,\, \alpha_3\simeq 0.1$ at the effective supersymmetry scale,
which we take to be 1 TeV. This solution is clearly unphysical.  
We are, therefore, led to set $\tilde Y_\tau^*=0$, yielding  the solution
\begin{eqnarray}
& \displaystyle \tilde Y_t^*={47 \tilde \alpha_1+225 \tilde \alpha_2
+200 \tilde \alpha_3\over 425}
\simeq 0.0052, & \\
& \displaystyle \tilde Y_b^*={13\tilde\alpha_1+225 \tilde \alpha_2
+200 \tilde \alpha_3\over 425}
\simeq 0.0052, &\\
& \displaystyle \tilde Y''^*={2\left(11\tilde \alpha_1-45 \tilde \alpha_2
+130 \tilde \alpha_3
\right)\over 255}
\simeq 0.0073, & 
\end{eqnarray}
which  leads to the quasi-fixed point values for these couplings
\begin{eqnarray}
& \displaystyle h_t^*\simeq 0.91, & \label{analyukawatop} \\
& \displaystyle h_b^*\simeq 0.90, & \label{analyukawab}\\
& \displaystyle \lambda^{''*}_{233}\simeq 1.08. & \label{analyukawalambda}
\end{eqnarray}
We note that these quasi-fixed point values for the respective
couplings  are 
similar to the values obtained for them in MSSM~\cite{ap2}.

\subsection{Quasi-fixed points for trilinear soft supersymmetry 
breaking parameters}
We now turn our attention to the renormalization group equations 
(\ref{rge7}) - (\ref{rge15}) for the $A$ parameters, and their
quasi-fixed points.  Since the quasi-fixed point values  
for $h_\tau, \lambda, \kappa$ are trivial,
we cannot determine the quasi-fixed point
values for $A_\tau, A_{\lambda}$, and $A_{\kappa}$,
and, therefore, ignore the coresponding renormalization group
equations~ (\ref{rge9}) - (\ref{rge11}). 
In rest of the $A$   equations, we substitute 
$h_\tau= \lambda = \kappa = 0$, and obtain
the equations that the remaining $A$ parameters must
satisfy in order for us to determine the quasi-fixed point solution:
\begin{eqnarray}
6 \tilde Y_t A_t + \tilde Y_b A_b + 2 \tilde Y'' A_{\lambda''}
-{16\over 3}\tilde \alpha_3 M_3
-3 \tilde \alpha_2 M_2 -{13\over 15} \tilde \alpha_1 = 0, & \\
\tilde Y_t A_t + 6 \tilde Y_b A_b + 2 \tilde Y'' A_{\lambda''}
-{16\over 3} \tilde \alpha_3 M_3
-3 \tilde \alpha_2 M_2 -{7\over 15} \tilde \alpha_1 = 0, & \\
2 \tilde Y_t A_t + 2 \tilde Y_b A_b + 6 \tilde Y'' A_{\lambda''}
-8\tilde \alpha_3 M_3
-{4\over 5} \tilde \alpha_1 = 0. & 
\end{eqnarray}
These yield the following quasi-fixed point solution:
\begin{eqnarray}
& \displaystyle A_t^*={47\tilde\alpha_1 M_1+225 \tilde\alpha_2 M_2
+200 \tilde\alpha_3 M_3
\over 425 \tilde Y_t^*}\simeq 0.77 m_{\tilde g},& \label{analtrilineartop} \\
& \displaystyle A_b^*={13\alpha_1 M_1+225 \tilde\alpha_2 M_2
+200 \tilde\alpha_3 M_3
\over 425 \tilde Y_b^*}\simeq 0.78 m_{\tilde g},& \label{analtrilinearb} \\
& \displaystyle A_{\lambda''}^*={2\left(11i\tilde\alpha_1 M_1
-45 \tilde\alpha_2 M_2
+130 \tilde\alpha_3 M_3\right)\over 255 \tilde Y''^*}
\simeq 1.02 m_{\tilde g}, & 
\label{analtrilinearlambda} 
\end{eqnarray}
where $m_{\tilde g}$ is the gluino mass~($ = M_3$) defined at the weak scale.
For the numerical estimates in
(\ref{analtrilineartop}) - (\ref{analtrilinearlambda}),
we have used the fact that the gaugino masses scale as the
square of the gauge couplings,  and that $\alpha_G\simeq 0.041$ 
at the grand unified scale 
$M_G\simeq 10^{16}$GeV.  One must compare these quasi-fixed point 
values with the true fixed-point values (\ref{truetrilinear}).
Since the quasi-fixed points for the Yukawa couplings represent
their upper bounds, the quasi-fixed point values (\ref{analtrilineartop}) - 
(\ref{analtrilinearlambda}) provide a lower bound on the
corresponding $A$ parameters. We note that inspite of the presence of 
additional  trilinear couplings in the superpotential of the NMSSM, 
the quasi-fixed point
behaviour is similar to the corresponding quasi-fixed point behaviour 
in MSSM with baryon and lepton number violation.

\section{Numerical Results and Discussion}

In the previous section we have obtained the approximate quasi-fixed point 
values for the Yukawa couplings and the $A$ parameters by an algebraic
solution of the corresponding RG equations.  The RG equations are a set
of coupled first-order  differential equations that must be solved
numerically to obtain accurate values for the fixed points. We have
numerically solved the RG equations for the Yukawa couplings, and the 
$A$ parameters. We now present the results of such a
numerical analysis.

In Fig.1 we show the fixed point behaviour of the top-quark Yukawa 
coupling as a function of the logarithm of the scale parameter
$\mu$. We have included the evolution equations for the b-quark,
the $\tau$-lepton Yukawa coupling, and the trilinear couplings
$\lambda$ and $\kappa$, as well as the
B-violating coupling $\lambda_{233}''$, and evolved them 
from the initial value $ln(\mu/M_G) = 0$ to the final
value  $ln(\mu/M_G) \simeq -33$ at the weak scale,  
for the numerical solution. 
It is seen that for all $h_t \stackrel{>}{\sim} 1$
at the GUT scale, the top-quark Yukawa coupling approaches its
quasi-fixed point at the weak scale. We note that the numerical
evolution of fixed point approaches but does not exactly reproduce
the approximate analytical value in (\ref{analyukawatop}). In Figs. 2 and 3
we present the corresponding approach to the quasi-infrared 
fixed point for 
the couplings $h_b$ and $\lambda_{233}''$, respectively. These infrared fixed
points provide a model independent 
theoretical upper bound on the $B$-violating
coupling $\lambda_{233}''$. It is worthwhile to point out here  that
the fixed point value for  $\lambda_{233}'' \simeq 1.08$ is more than a 
factor of $3$ lower than 
what would  be a naive perturbative upper bound of 
$\lambda_{233}''\stackrel{<}{\sim} \sqrt 4\pi \simeq 3.5$
for this coupling.

In Figs. 4, 5 and 6, we present the fixed point behaviour of the 
corresponding $A$ parameters. We notice the remarkable focussing
property seen in the fixed point behaviour of all the $A$ parameters.
Again, we notice that the numerical evolution of the fixed point
approaches, but does not actually reproduce, the approximate analytical
values of Eqs.~(\ref{analtrilineartop}), (\ref{analtrilinearb}),  and
(\ref{analtrilinearlambda}).  Since the quasi-fixed
point value for the $A$ parameter is inversely
proportional to the quasi-fixed point value of the Yukawa coupling,
it provides a lower bound on the corresponding 
$A$ parameter.

\section{Summary and Conclusions}

We have carried out a comprehensive study of the
renormalization group flow in the nonminimal supersymmetric
standard model
with all the third generation Yukawa couplings, and the 
trilinear couplings $\lambda$ and $\kappa$, and with
highest generation
baryon and lepton number violation. We have shown that the simultaneous
fixed point for the top- and bottom-Yukawa couplings, and the $B$-violating
coupling $\lambda_{233}''$, is the only fixed point that is
stable in the infra-red region. However, the top-quark mass
predicted by this fixed point is incompatible with measured value of the 
top mass. This fixed point, therefore, provides a process-independent
lower bound on the baryon number violating coupling $\lambda''_{233}$
at the electroweak scale.

We have shown that all other possible fixed point  solutions are 
either unphysical,  or unstable, in the infra-red region. 
{\it In particular
there is no infrared fixed point with simultaneous $B$ and $L$ violation.}
This result is analogous to that found in MSSM with $B$ and
$L$ violation~\cite{ap2}.

We have also carried out the renormalization group analysis of the 
corresponding trilinear soft supersymmetry breaking parameters.
We have obtained the true fixed points for these parameters,
which serve as  upper bounds on these parameters.

Since the true fixed points are not reached in practice at the 
electroweak scale, we have also obtained the quasi-fixed points of the 
Yukawa couplings and the trilinear parameters. The quasi-fixed point
values for the Yukawa couplings are numerically 
very close to the values obtained
previously by ignoring the $\tau$ Yukawa coupling. 
Since the quasi-fixed points are  reached for large initial
values of the couplings at the GUT scale, these reflect on the
assumption of  perturbative unitarity, or the
absence of Landau poles, of the corresponding couplings.
These quasi-fixed points, therefore,   provide an upper bound on the 
relevant Yukawa coupling, especially the baryon number violating
coupling $\lambda_{233}''$.  From the true fixed point and the quasi-fixed
point ananlysis we are able to constrain the baryon number violating
coupling $0.97 \stackrel{<}{\sim} \lambda_{233}'' \stackrel{<}{\sim} 
1.08$ in a model independent manner.
We would like to emphasize that this infrared fixed point upper bound
on  $\lambda_{233}''$ is more than a factor of 3 lower than 
the naive perturbative upper bound on this coupling, and
represents a significant constraint on it.
We have complemented the quasi-fixed point analysis of the Yukawa couplings 
with an  analysis of the corresponding soft supersymmetry
breaking trilinear couplings. We have shown that the $A$ parameters
for the top- and bottom-quark Yukawa couplings, and the baryon number
violating couplings all show  striking convergence properties.
This strong focussing property is quite independent of
the input parameters at the
unification scale (or equivalently the pattern of supersymmetry breaking),
and the $A$ parameters are, therefore, fully determined
in the quasi-fixed regime. 
In particular, we have constrained  the $A$ parameters to be
$0.77 \stackrel{<}{\sim} A_t/m_{\tilde g} \stackrel{<}{\sim} 1$,
$0.78 \stackrel{<}{\sim} A_b/m_{\tilde g} \stackrel{<}{\sim} 1$,
and $A_{\lambda''}/m_{\tilde g} \simeq 1$.
These constraints are analogous to those found in MSSM for these 
parameters. Thus, the infrared fixed point behaviour 
of NMSSM with $B$ and $L$ violation is similar to that of MSSM, 
and may be universal for supersymmetric models based on 
$SU(2)_L \times U(1)_Y$ gauge group.

\newpage

\section{\bf Acknowledgements}

The author is supported by  the University Grants Commission
under project No. 30-63/98/SA-III. He would like to thank 
the Theory Group at DESY for its hospitality 
while this work was completed.

\newpage

\noindent{\bf Figure Captions}

\bigskip

\noindent {\bf Fig. 1.}  The quasi-fixed point 
of the renormalization group evolution of the
top-quark Yukawa coupling $h_t$ as a function of the logarithm
of the energy scale. We have taken the initial values of
$h_t$ at the grand unified scale
$M_G \sim 10^{16}$ to be $4.0, \, 3.0, \,
2.0$, and $1.0$, and evolved them down to
$ln(\mu/M_G) \simeq -33$.  The initial values of other Yukawa couplings 
are $h_b = 1.0,\, h_\tau=0,\, \lambda = \kappa = 0.5$  
and $\lambda''_{233}= 2.0$.

\medskip

\noindent {\bf Fig. 2.} The renormalization group evolution of the  
bottom-quark Yukawa coupling $h_b$ as a function of the logarithm
of the energy scale. The initial values 
of $h_b$ at $M_G$ are $4.0, \, 3.0, \, 2.0,$ and $1.0.$
The initial values of  other Yukawa couplings are 
$h_t=1.0,\, h_\tau=0,\, \lambda = \kappa = 0.5, \, 
\lambda''_{233}=2.0$. 

\medskip

\noindent {\bf Fig. 3.}  Th equasi-fixed point behaviour of the
baryon number violating Yukawa coupling $\lambda''_{233}$ 
as a function of the logarithm of the energy scale.
The initial values are $\lambda''_{233} = 4.0, \, 3.0, \, 2.0, \, 1.0$.
The initial values of other other parameters are 
are $h_t=h_b=1.0,\, h_\tau=0, \, \lambda = \kappa = 0.5$.

\medskip

\noindent {\bf Fig. 4.}  Renormalization group evolution of ratio 
$A_t/m_{\tilde g}$ as a function of the logarithm of the energy scale for 
several different initial values at the grand unified scale $M_G$.
The initial values at $M_G$ are $-4, -3, -2, -1, 0, 1, 2, 3, 4.$
The initial values for other parameters at $M_G$ are
$h_t=4.0, \, h_b=1.0,\, h_\tau=0,\, \lambda = \kappa = 0.1,
\lambda''_{233}=1.0$,
and $A_b/m_{\tilde g} = 2.0, \, A_{\lambda_{233}''}/m_{\tilde g} = 3.0$.

\medskip

\noindent {\bf Fig. 5.}  Renormalization group evolution of the
ratio $A_b/m_{\tilde g}$ as a function 
of the logarithm of the energy scale.
The initial values of the ratio are
$-4, -3, -2, -1, 0, 1, 2, 3, 4.$
Other parameters at the scale $M_G$ are
$h_t=1.0,\, h_b = 4.0, \, h_\tau=0,\, \lambda = \kappa = 0.1, \,
\lambda''_{233}=1.0$,
and $A_t/m_{\tilde g}= 2.0, \, A_{\lambda_{233}''}/m_{\tilde g}=2.57$.

\medskip

\noindent {\bf Fig. 6.}  Renormalization group evolution of the
trilinear coupling $A_{\lambda_{233}''}/m_{\tilde g}$.
The initial values of the ratio are
$-4, -3, -2, -1, 0, 1, 2, 3, 4.$
Other parameters at the scale $M_G$ are
$h_t=h_b=1.0, \, h_\tau=0, \, \lambda = \kappa = 0.1$,
and $A_t/m_{\tilde g} = 2.0 \, A_b/m_{\tilde g} = 2.0$.

\begin{center}
\begin{figure}\nonumber
\epsfig{figure=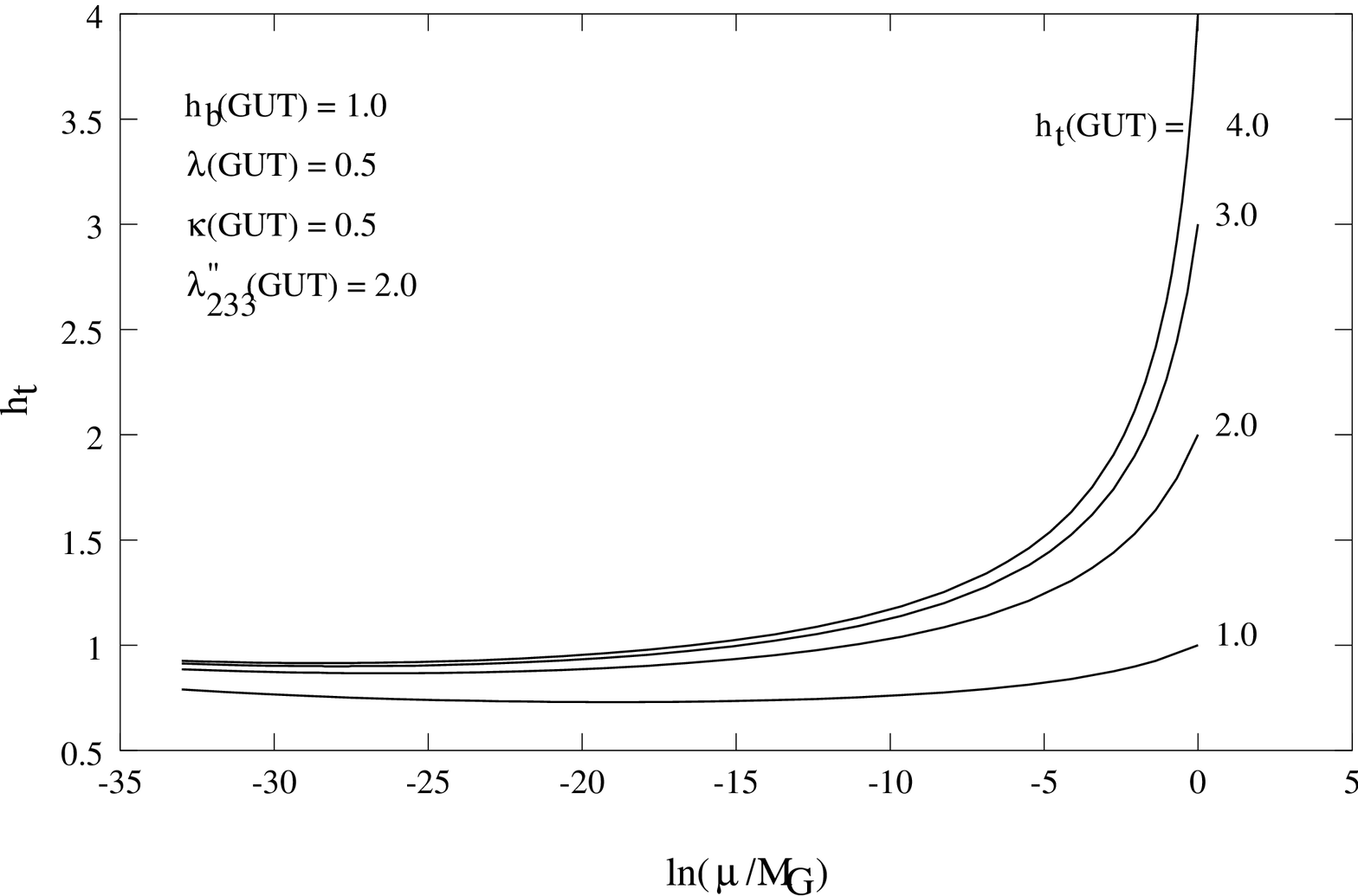,width=15cm,height=15cm}
\vspace{2cm}
\caption{}
\end{figure}

\begin{figure}\nonumber
\epsfig{figure=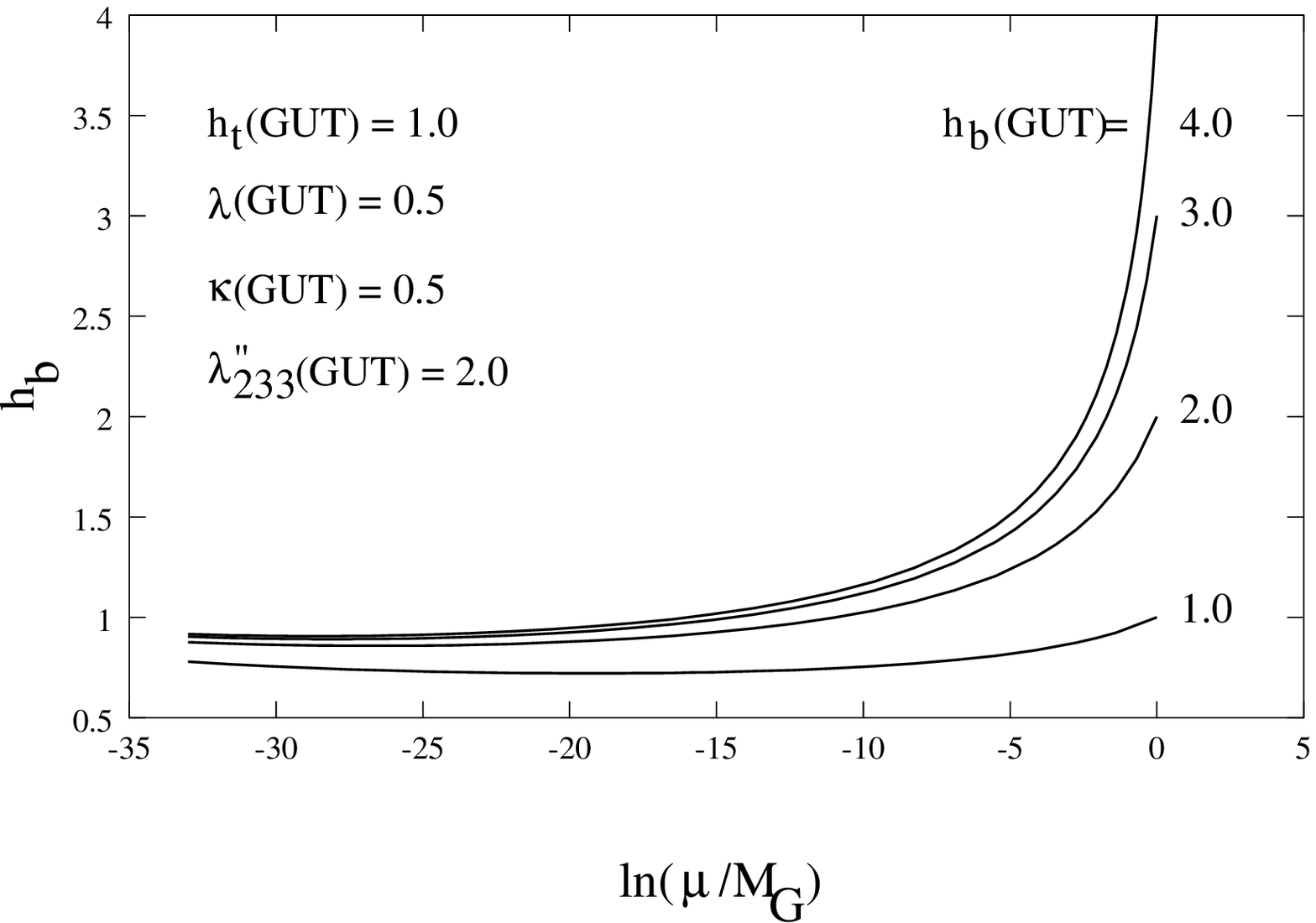,width=15cm,height=15cm}
\vspace{2cm}
\caption{}
\end{figure}

\newpage

\begin{figure}\nonumber
\epsfig{figure=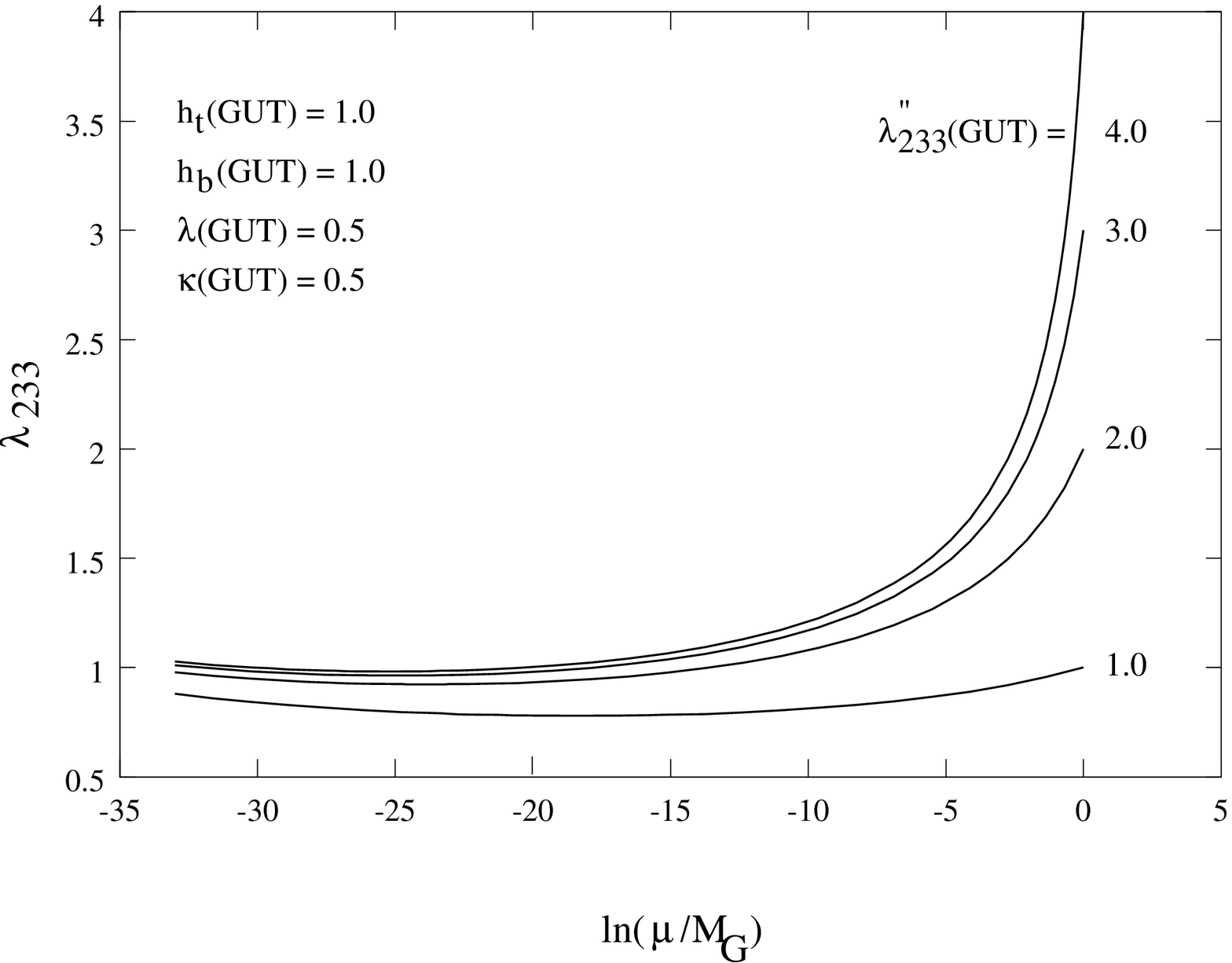,width=15cm,height=15cm}
\vspace{2cm}
\caption{}
\end{figure}

\begin{figure}\nonumber
\epsfig{figure=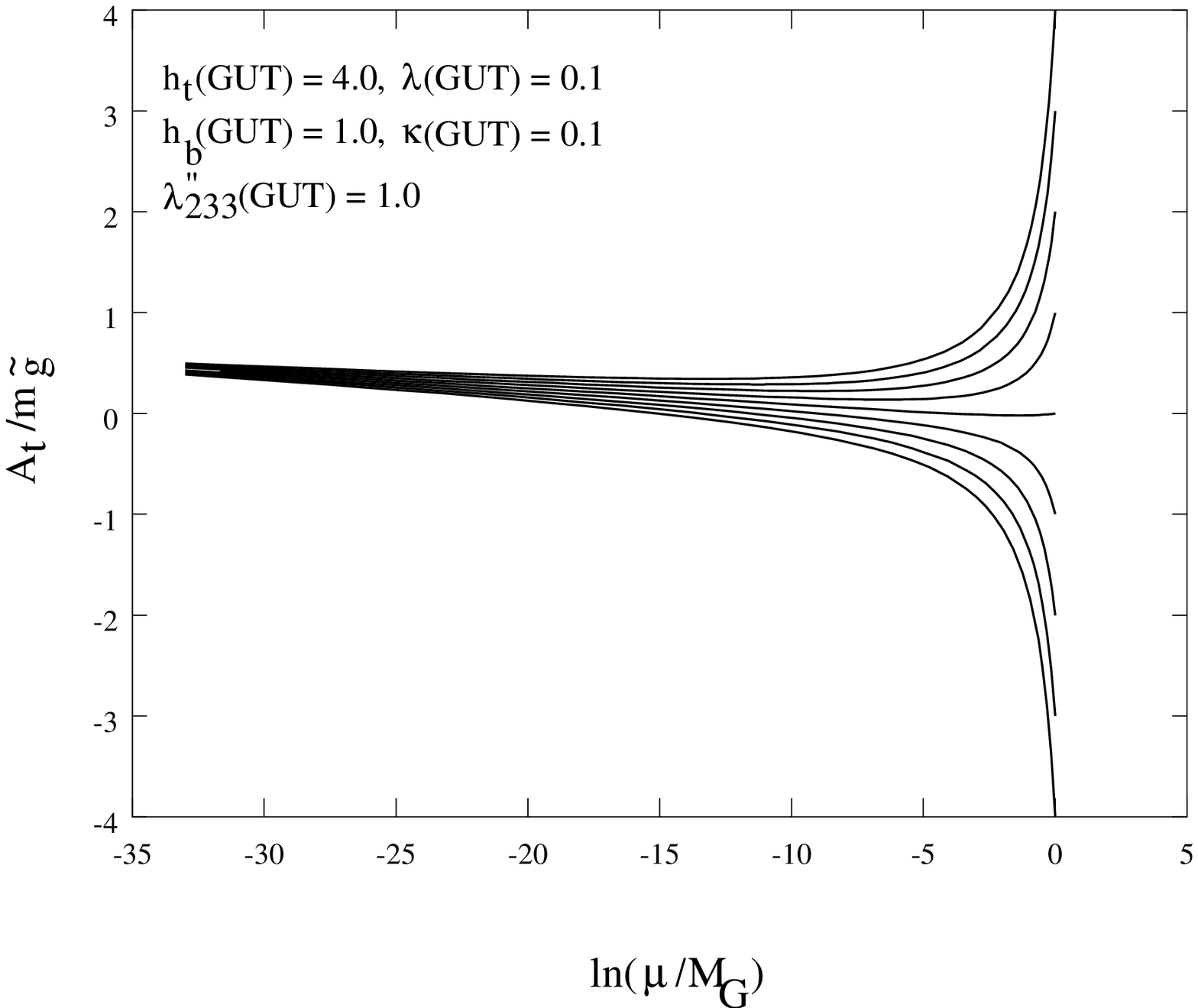,width=15cm,height=15cm}
\vspace{2cm}
\caption{}
\end{figure}

\begin{figure}\nonumber
\epsfig{figure=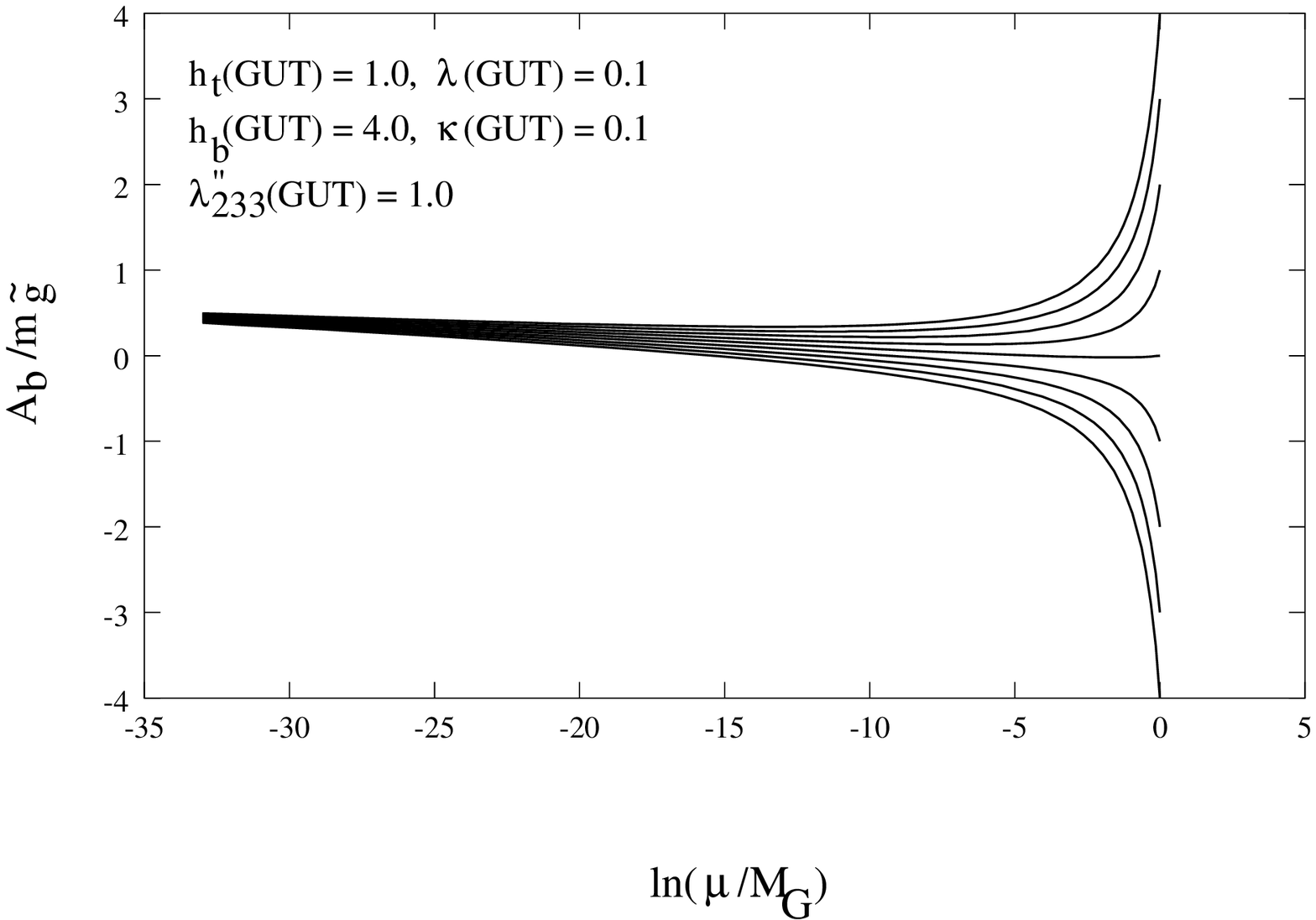,width=15cm,height=15cm}
\vspace{2cm}
\caption{}
\end{figure}

\begin{figure}\nonumber
\epsfig{figure=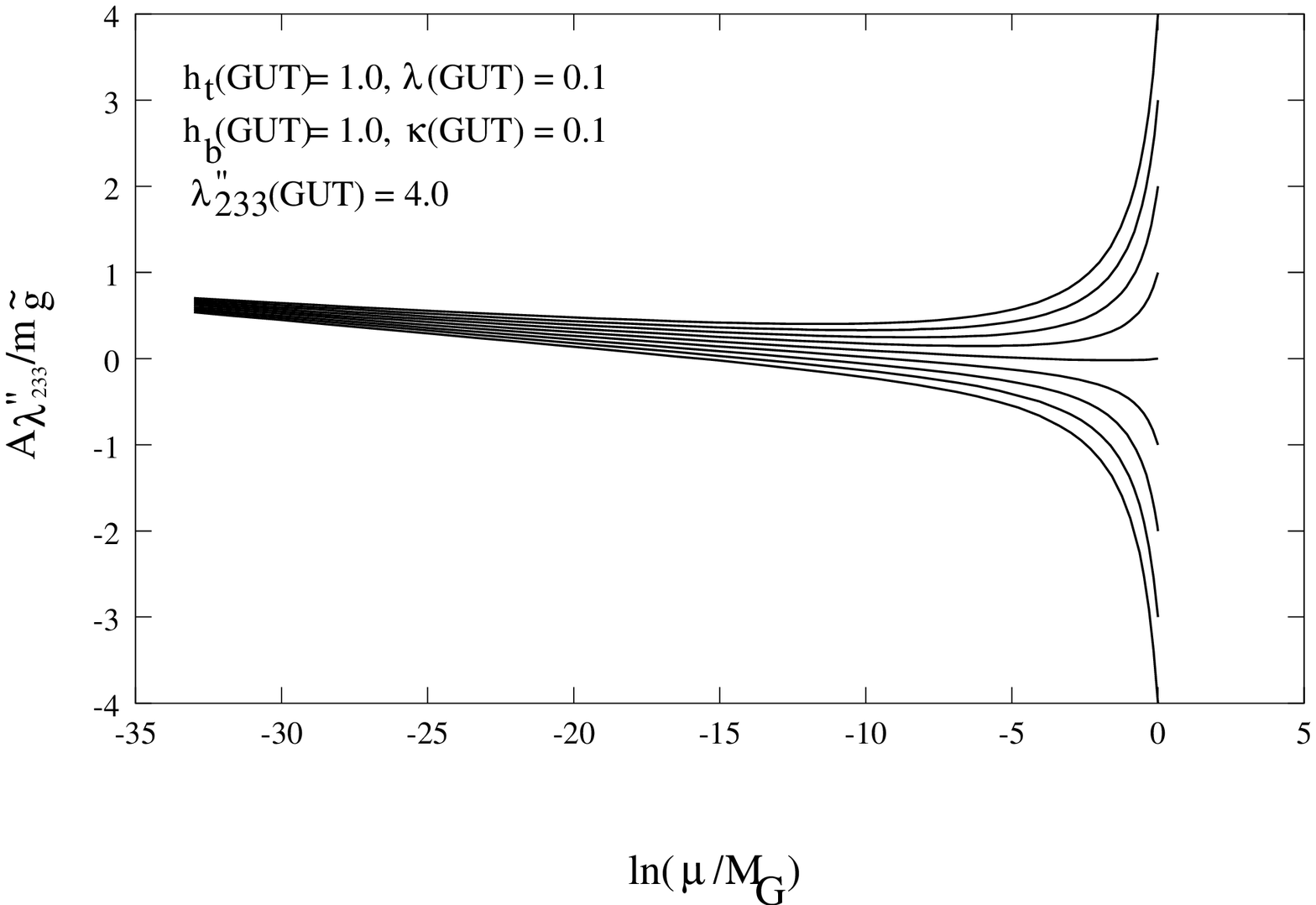,width=15cm,height=15cm}
\vspace{2cm}
\caption{}
\end{figure}

\end{center}


\begin{thebibliography}{abcdef}

\bibitem{wess} J. Wess and J. Bagger, ``Supersymmetry and Supergravity''
(Princeton University Press, Princeton, NJ, 1992);
P. Nath, R. Arnowitt and A. H. Chamseddine, `` Applied N = 1 Supergravity''
(World Scientific, Singapore, 1984).


\bibitem{schrempp1} For a review and references, see e.g.,
B. Schrempp and M. Wimmer, Prog. Part. Nucl. Phys.
{\bf 37},  1 (1996).

\bibitem{pendross} B. Pendleton and G. G. Ross, Phys. Lett. {\bf B98},
291 (1981); M. Lanzagorta and G. G. Ross, Phys. Lett. {\bf B349}, 
319 (1995).

\bibitem{hill} C. T. Hill, Phys. Rev.  {\bf D24}, 691 (1981).

\bibitem{allanach} B. C. Allanach and S. F. King, Phys. Lett. 
{\bf B407}, 124 (1997).

\bibitem{abel} S. A. Abel and B. C. Allanach, Phys. Lett.  {\bf B415},
371 (1997).

\bibitem{jack} I. Jack and D. R. T. Jones, Phys. Lett.  {\bf B443},
177 (1998).


\bibitem{weinberg} S. Weinberg, Phys. Rev. {\bf D26}, 287 (1982);
N. Sakai and T. Yanagida, Nucl. Phys. {\bf B197}, 133 (1982).

\bibitem{farrar} G. Farrar and P. Fayet, Phys. Lett. {\bf B76},  575 (1978).

\bibitem{ap1} B. Ananthanarayan and P. N. Pandita, Phys. Lett. 
{\bf B454}, 84 (1999).

\bibitem{ap2} B. Ananthanarayan and P. N. Pandita, Phys. Rev {\bf D62},
036009 (2000), and references therein; B. Ananthanarayan and
P. N. Pandita, Phys. Rev. {\bf D62}, 0760XX (2001) (in press).

\bibitem{fayet} P. Fayet, Nucl. Phys.  {\bf B90}, 104 (1975);
R. K. Kaul and P. Majumdar, Nucl. Phys. {\bf B199}, 36 (1982);
J. Ellis et al., Phys. Rev. {\bf D39}, 844 (1989);
P. N. Pandita, Phys. Lett. {\bf B318}, 338 (1993); Z. Phys. {\bf C59},
575 (1993).

\bibitem{paulraj} P. N. Pandita and P. Francis Paulraj,
Phys. Lett. {\bf B462}, 294  (1999).

\bibitem{barbier} See, e.g., R. Barbier et al., hep-ph/9810232 (unpublished);
B. Allanach et al., hep-ph/9906224.

\bibitem{nfal} N. Falck, Z. Phys. {\bf C30}, 247 (1986);
S.P. Martin and M.T. Vaughn, Phys. Rev. {\bf D50}, 2282 (1994).

\bibitem{PP21} These equations are generalizations of the
RG equations in P.N. Pandita and P. Francis Paulraj, Phys. Lett. 
{\bf B462}, 294 (1999) when we consider simultaneous highest
generation baryon and lepton number violating couplings in NMSSM.


\bibitem{codkaz1} S. Codoban and D. I. Kazakov, Eur. Phys. J. {\bf C13},
671 (2000).

\bibitem{rges} B. Gato et al., Nucl. Phys.  {\bf B253}, 285 (1985);
N. K. Falck, Z. Phys. {\bf C30}, 247 (1986);  S. P. Martin and
M. T. Vaughn, Phys. Rev. {\bf D50}, 2282 (1994).


\bibitem{pdg} Particle Data Group, D. E. Groom et al., 
Eur. Phys. J.  {\bf C},
1 (2000).

\bibitem{auberson1}
We follow here the notation of B. Ananthanarayan and
P. N. Pandita, Phys. Rev. {\bf D62}, 0760XX (2001) (in press).
See also G.~Auberson and G.~Moultaka,
Eur.\ Phys.\ J.\  {\bf C12}, 331 (2000);
D. Kazakov and G. Moultaka, Nucl. Phys. {\bf B577}, 121 (2000).

\bibitem{barger}
V. Barger, M. S. Berger, R. J. N. Phillips, and
T. W{\"o}hrmann, Phys. Rev. {\bf D53}, 6407 (1996) 



\end{thebibliography}
\end{document}